\documentclass[twocolumn,superscriptaddress,preprintnumbers,amsmath,amssymb,prc,nofootinbib]{revtex4-2}

\usepackage{graphicx}
\usepackage{dcolumn}
\usepackage{bm}
\usepackage{ifpdf}
\usepackage{epstopdf}
\usepackage{slashed}
\usepackage{amsfonts}
\usepackage{mathrsfs}
\usepackage{amssymb}
\usepackage{multirow}
\usepackage{CJK}
\usepackage{xcolor}
\usepackage{ulem}

\def\lesssim{\ \raise.3ex\hbox{$<$}\kern-0.8em\lower.7ex\hbox{$\sim$}\ }
\def\gesim{\ \raise.3ex\hbox{$>$}\kern-0.8em\lower.7ex\hbox{$\sim$}\ }

\begin{document}
\begin{CJK}{UTF8}{ipxm}

\title{Mass-imbalance effect on the cluster formation in a one-dimensional Fermi gas with coexistent $s$- and $p$-wave interactions}

\author{Yixin Guo (郭一昕)}
\email{yixin.guo@riken.jp}
\affiliation{RIKEN Nishina Center for Accelerator-Based Science, Wako 351-0198, Japan}

\date{\today}

\begin{abstract}
We consider the mass-imbalance effect on the clustering in a one-dimensional two-component Fermi gas with coexistent even- and odd-wave interactions resulting in different configurations of clustering phases.
We obtain the solutions of both stable two- and three-body cluster states with different mass ratios and configurations by solving the corresponding variational equations.
We numerically map out phase diagrams consisting of the $s$- and $p$-wave pairing phases, and {trimer} phase with different configurations, in a plane of $s$- and $p$-wave pairing strengths.
{Within the explored ranges of $s$- and $p$-wave pairing strengths, the in-vacuum three-body states are always more deeply bound than the two-body ones.}
While for the in-medium case, the Cooper {trimer} phase dominates over the pairing phases when both $s$- and $p$-wave interactions are moderately strong.
There is also a competition between different clustering configurations of three-body clustering.
\end{abstract}

\maketitle

\section{Introduction}\label{sec:I}

The investigation of superconductors and superfluids has been one of the frontiers in modern physics.
It is indicated that the competition and the coexistence of various orders are worth being investigated~\cite{Yasui2020PhysRevC.101.055806,marques2021quest,duer2022observation,Kanasugi2022Comm.Phys.5.1.,Guo2023Phys.Rev.B107.024511}, which have important applications in different fields such as condensed-matter and nuclear physics.
However, to understand the microscopic mechanism of non-trivial states arising in the mass-imbalanced systems with coexistent orders is still one of the most challenging open questions.

It would be useful to study the mass-imbalance effect on multibody clustering in various systems.
The Efimov state~\cite{Efimov1970Phys.Lett.B33.563--564} has attracted lots of interest in ultracold atoms, nuclear physics, and so on.
In the corresponding investigation, experimental evidence of Efimov physics has been reported in different mass-imbalanced ultracold atomic gas mixtures such as $^{41}$K-$^{87}$Rb~\cite{Barontini2009Phys.Rev.Lett.103.043201} or $^6$Li-$^{133}$Cs~\cite{PhysRevLett.112.250404} ones.
The two-nucleon halo nuclei such as $^{11}$Li ($^9$Li core $+$ two halo neutrons) is also regarded as an ideal candidate to experimentally detect the Efimov states~\cite{PhysRevLett.73.2817}.
So far, various possible superconductors with competing odd- and even-parity pairing channels have been experimentally realized in the mass-imbalanced systems, including UTe$_2$~\cite{Kanasugi2022Comm.Phys.5.1.,Jiao2020Nature579.523--527}, UPt$_3$~\cite{Sauls1994Adv.Phys.43.113--141}, WTe$_2$~\cite{Jia2022Nat.Phys.18.87--93}, and Sr$_2$RuO$_4$~\cite{Mackenzie2003Rev.Mod.Phys.75.657--712,Kinjo2022Science376.397--400}.
Moreover, in nuclear physics, there is also a novel state called hypernucleus consisting of at least one hyperon in addition to the normal nucleons~\cite{Danysz1953}.
It shows many interesting properties, such as that hypernuclei containing the lightest hyperon $\Lambda$ (such as $^7_\Lambda$Li) tend to be more tightly bound than normal nuclei~\cite{PhysRevLett.86.1982}.
The hypernuclei with small particle number would be useful for the preliminary studies.

The ultracold atomic gas is regarded as an ideal platform for investigating such non-trivial quantum many-body phenomena, due to its tunable interactions and densities~\cite{Chin2010Rev.Mod.Phys.82.1225--1286,Ohashi2020Prog.Part.Nucl.Phys.111.103739}.
Both $s$- and $p$-wave interactions are tunable near the Feshbach resonance~\cite{Ticknor2004Phys.Rev.A69.042712,Gurarie2005Phys.Rev.Lett.94.230403,Schunck2005Phys.Rev.A71.045601,Inada2008PhysRevLett.101.100401,Ohashi2020Prog.Part.Nucl.Phys.111.103739,Nakasuji2013PhysRevA.88.012710,Strinati2018Phys.Rep.738.1--76}.
Consequently, it is interesting to realize the situation that Fermi superfluids with hybridized $s$- and $p$-wave pairings in cold atomic systems via overlapped Feshbach resonances~\cite{Chin2010Rev.Mod.Phys.82.1225--1286,Regal2003PhysRevLett.90.053201,Zhou2017ScienceChina.60.12}.
In such a system, in addition to $s$- and $p$-wave Cooper pairings, the clustering associated with the Cooper instability may be extended to induce the bound states consisting of three particles or even more.
In order to study such a many-body problem, the generalized Cooper problem has been further applied to cluster states such as three-body~\cite{Niemann2012Phys.Rev.A86.013628,Kirk2017Phys.Rev.A96.053614,Tajima2021Phys.Rev.A104.053328,Tajima2022Phys.Rev.Research4.L012021, Guo2022Phys.Rev.A106.043310,Guo2023Phys.Rev.B107.024511,Guo2023Phys.Rev.A108.043303} and even four-body Cooper bound states~\cite{Roepke1998Phys.Rev.Lett.80.3177--3180,Sandulescu2012Phys.Rev.C85.061303,Baran2020Phys.Lett.B805.135462,Guo2022Phys.Rev.C105.024317,Guo2022Phys.Rev.Research4.023152,Guo2025Phys.Rev.C112.024310}.
On the other hand, with the help of specific geometry and trap potential, low dimensions and lattice structures can be achieved in cold-atomic systems~\cite{Bloch2008Rev.Mod.Phys.80.885--964,PhysRevLett.94.210401,Liao2010Nature567,Sowiski2019}.
Such low-dimensional systems in condensed-matter or nuclear physics have also attracted much attention~\cite{Alexandrou1989PhysRevC.39.1076,Hagino2010,altomare2013one}.

\begin{figure}
  \includegraphics[width=1.0\linewidth]{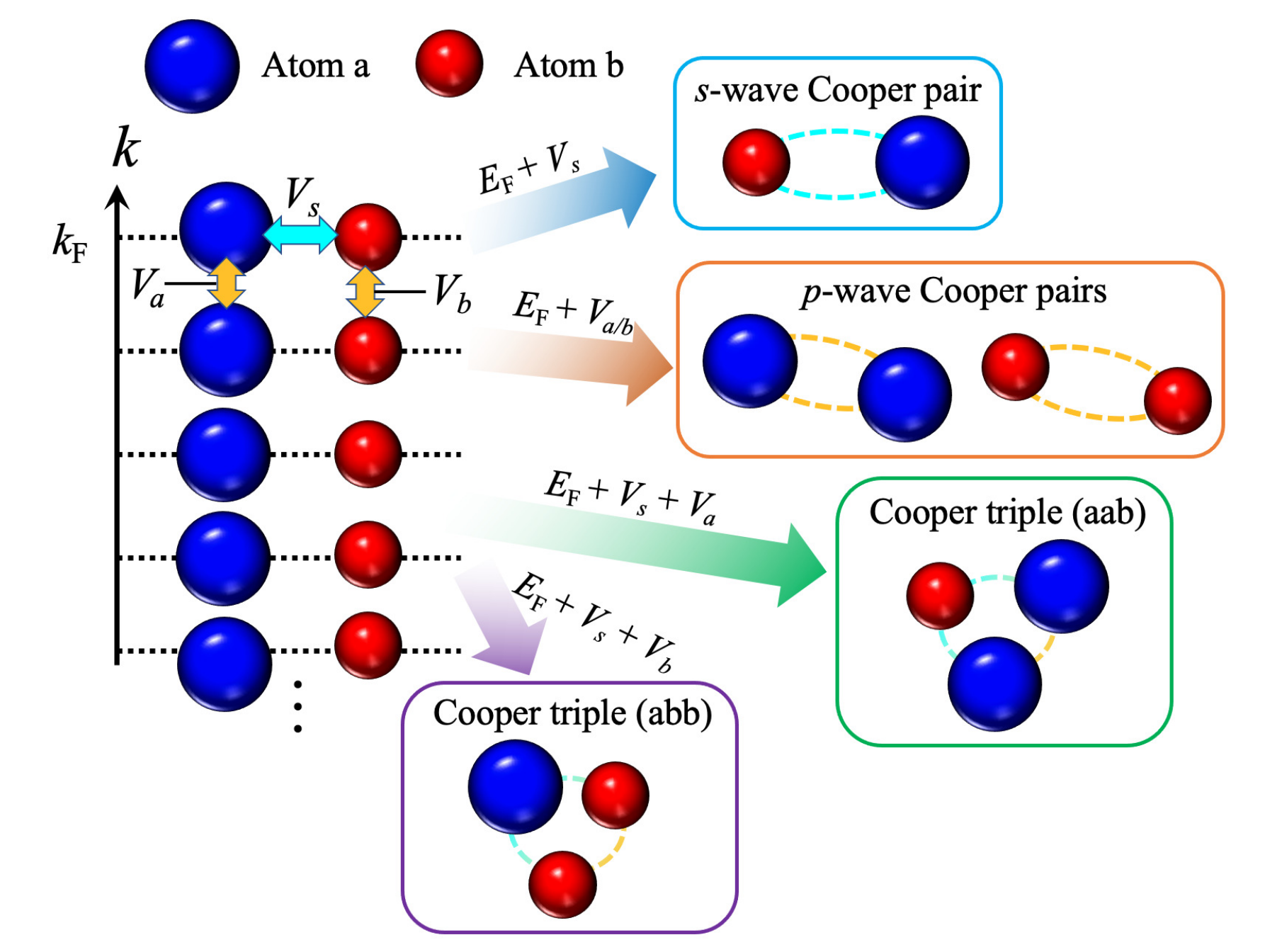}
  \caption{Schematic figure representing our model.
  We consider the degenerate two-component mass-imbalanced fermions (states $a$ and $b$ occupying the levels with momentum $k$ up to the Fermi momentum $k_{\rm F}$) and the consequence of coexistent interspecies $s$-wave interaction $V_s$ and intraspecies $p$-wave interactions $V_i$ (acting on two identical fermions in the $i=a,b$ state),
  which lead to the Cooper instabilities towards the $s$- and $p$-wave Cooper pairs, and the Cooper {trimers}.}\label{fig:0}
\end{figure}

In the previous work~\cite{Guo2023Phys.Rev.B107.024511}, we studied the degenerate two-component fermions with coexistent interspecies and intraspecies interactions, which lead to the Cooper instabilities towards the $s$- and $p$-wave Cooper pairs, and the three-body Cooper bound states with mass-balanced configuration.
{In the following, we refer to the formation of three-body Cooper bound states as Cooper trimer formation.}
However, the mass imbalance is also worth being investigated, and the intraspecies interaction is not necessarily limited to exist only within one component.
In this work, we are going to figure out the properties in a system with both $s$- and $p$-wave interactions as shown in Fig.~\ref{fig:0}.
The two- and three-body clusterings with different configurations are also anticipated.
In Ref.~\cite{Jackson2023Phys.Rev.X13.021013}, by activating the orbitals degree of freedom, the authors reported that the emergence of even-parity ($s$-wave) interaction in a spin-polarized fermionic $^{40}$K gas besides the odd-wave ($p$-wave) interactions among identical fermions in each orbital.
Based on that, present model setup might be achieved experimentally via fulfilling multiorbital structure with different components of atoms in the future.

In this paper, we investigate the fate of in-vacuum and in-medium two- and three-body correlations in a one-dimensional two-component Fermi gas with coexistent $s$- and $p$-wave interactions.
Specific attention is paid to the mass imbalance and different configurations of clustering.
In general, the mass of two components of fermions are adopted as different ones, and $p$-wave interactions among both kinds of atoms are taken into account.
By solving the variational equations, we show the solutions of the stable clusters.
We also figure out a phase diagram consisting of pairing and {trimer} states with different configurations.

This paper is organized as follows.
The theoretical framework is presented in Sec.~\ref{sec:II}, where we show the Hamiltonian for present system. 
We apply a variational approach for in-medium two- and three-body states on top of the Fermi sea to this model. 
In Sec.~\ref{sec:IIIA}, we first discuss our numerical results for the in-vacuum bound states and the {lowest-energy cluster-instability diagram}.
The results of in-medium case are shown in Sec.~\ref{sec:IIIB}.
Finally, a summary and perspectives will be given in Sec.~\ref{sec:IV}.
In the following, we take $\hbar=c=k_{\rm B}=1$.
The system size is taken to be a unit.

\section{Theoretical framework}\label{sec:II}

In this work, we consider a one-dimensional mass-imbalanced two-component fermions with coexistent short-range interactions.
The intraspecies odd-parity ($p$-wave) interactions among each kind of fermion are both taken into {account}.
{As for} the interspecies interaction, we assume the even-parity ($s$-wave) one is dominant and neglect the odd-parity ($p$-wave) interaction one as the same as the previous work~\cite{Guo2023Phys.Rev.B107.024511}.
{We note that in a two-component system this assumption does not restrict the possible three-body clustering configurations, and an interspecies odd-parity interaction would only introduce higher-order contributions within the same configurations.}
Consequently, the present system can be described by the following Hamiltonian as
\begin{align}
    H=\,&K+V_s+V_{a}+V_{b},\\
    K=\,&\sum_{k}\left(\xi_{k,a}a_{k}^\dag a_{k}+\xi_{k,b}b_{k}^\dag b_{k}\right),\\
    V_{a}=\,&\frac{U_{a}}{2}\sum_{p,p',q}pp'a_{p+q/2}^\dag a_{-p+q/2}^\dag a_{-p'+q/2} a_{p'+q/2},\\
    V_{b}=\,&\frac{U_{b}}{2}\sum_{p,p',q}pp'b_{p+q/2}^\dag b_{-p+q/2}^\dag b_{-p'+q/2} b_{p'+q/2},\\
    V_s=\,&U_s\sum_{p,p',q}
    a_{
     p+ \frac{m_r}{m_b}q}^\dag 
    b_{-
     p+\frac{m_r}{m_a}q}^\dag 
    b_{-
     p'+ \frac{m_r}{m_a}q} 
    a_{
     p'+ \frac{m_r}{m_b}q},
\end{align}
where $\xi_{k,i}=k^2/(2m_i)-\mu_i$ ($i=a,b$) in the kinetic term $K$ is the single-particle energy with a momentum $k$, an atomic mass $m_i$, and chemical potential $\mu_i$.
Here the reduced mass reads $m_r=\frac{m_am_b}{m_a+m_b}$.
Note that in general the cases with $m_a\neq m_b$ will be considered in this paper.
In the generalized Cooper problems, we take $\mu_i=E_{\textrm{F},i}$ where $E_{\textrm{F},i}$ is the Fermi energy of species $i$.
{Namely, throughout the parameter scans, the chemical potentials are fixed by the Fermi energies $\mu_i=k_{\rm{F}}^2/(2m_i)$, corresponding to equal densities but different masses.}
Due to the mass difference, there is a mismatch of the Fermi energy (or chemical potential) between two different kinds of fermions, namely, $\mu_a\neq\mu_b$.
$V_i$ represents the two-body $p$-wave interaction between two identical atoms $i$ with a coupling constant $U_i$, and $V_s$ corresponds to the two-body $s$-wave interaction with a coupling constant $U_s$.
The contact couplings $U_s$ and $U_i$ can be renormalized by introducing the $s$-wave and $p$-wave scattering lengths as
\begin{align}
    U_s=-\frac{1}{m_ra_s},
\end{align}
and
\begin{align}
    \frac{m_i}{2a_{p}}=\frac{1}{U_i}+\sum_{p}\frac{p^2}{2\varepsilon_{p,i}}
\end{align}
with $\varepsilon_{p,i}=p^2/(2m_i)$, respectively.
Here we assume the $p$-wave scattering lengths of two components are always tuned to be equal to each other.
Incidentally, we are interested in an attractive $s$-wave interaction in this work, the positive $s$-wave scattering length $a_s>0$ is taken.

As schematically shown in Fig.~\ref{fig:0}, with the mass imbalance and the $p$-wave interactions among both kinds of atoms, there are in total five phases of correlations expected to appear in the present system.
In detail, as for the pairing states, we have $s$-wave pairing, $p$-wave pairing ($aa$  configuration) and $p$-wave pairing ($bb$ configuration).
While for the {trimer} states, we have $aab$ and $abb$ configurations, respectively.

The trial wave functions for these in-medium two-and three-body states are adopted as
\begin{align}\label{eq8}
    |\Psi_{ab}\rangle
    =\,&\sum_{k}
    \theta(|k|-k_{\rm F})
    \Omega^{ab}_{k}a_{k}^\dag b_{-k}^\dag|{\rm FS}\rangle\cr
    =\,& \sum_{k}'\Omega^{ab}_{k} a_{k}^\dag b_{-k}^\dag|{\rm FS}\rangle,
\end{align}
for $s$-wave pairing state~\cite{Ohashi2020Prog.Part.Nucl.Phys.111.103739},
\begin{align}
    |\Psi_{aa}\rangle=\,&\sum_{k}
    \theta(|k|-k_{\rm F})
    \Omega^{aa}_{k}a_{k}^\dag a_{-k}^\dag|{\rm FS}\rangle\cr
    =\,& \sum_{k}'\Omega^{aa}_{k}a_{k}^\dag a_{-k}^\dag|{\rm FS}\rangle,
\end{align}
for $p$-wave pairing ($aa$  configuration) state~\cite{Guo2022Phys.Rev.A106.043310},
\begin{align}
    |\Psi_{bb}\rangle=\,&\sum_{k}
    \theta(|k|-k_{\rm F})
    \Omega^{bb}_{k}b_{k}^\dag b_{-k}^\dag|{\rm FS}\rangle\cr
    =\,& \sum_{k}'\Omega^{bb}_{k} b_{k}^\dag b_{-k}^\dag|{\rm FS}\rangle,
\end{align}
for $p$-wave pairing ($bb$ configuration) state,
\begin{align}
    |\Psi_{aab}\rangle=\,&\sum_{p,q}
    \theta(|p+q/2|-k_{\rm F})\theta(|-p+q/2|-k_{\rm F})\nonumber\\
   & \times\theta(|-q|-k_{\rm F})
    \Omega^{aab}_{p,q}a_{p+q/2}^\dag a_{-p+q/2}^\dag b_{-q}^\dag|{\rm FS}\rangle\cr
    =\,& \sum_{p,q}'\Omega^{aab}_{p,q}a_{p+q/2}^\dag a_{-p+q/2}^\dag b_{-q}^\dag|{\rm FS}\rangle,
\end{align}
for {Cooper trimer} ($aab$ configuration) state~\cite{Guo2023Phys.Rev.B107.024511}, and
\begin{align}\label{eq12}
    |\Psi_{abb}\rangle=\,&\sum_{p,q}
    \theta(|p+q/2|-k_{\rm F})\theta(|-p+q/2|-k_{\rm F})\nonumber\\
   & \times\theta(|-q|-k_{\rm F})
    \Omega^{abb}_{p,q}a_{-q}^\dag b_{p+q/2}^\dag b_{-p+q/2}^\dag|{\rm FS}\rangle\cr
    =\,& \sum_{p,q}'\Omega^{abb}_{p,q}  a_{-q}^\dag b_{p+q/2}^\dag b_{-p+q/2}^\dag|{\rm FS}\rangle,
\end{align}
for {Cooper trimer} ($abb$ configuration) state, respectively.
In the above trial wave functions, $\Omega^{mn\cdots}_{q_1,q_2,\cdots}$ denotes the variational parameter for the state with $mn\cdots$ configuration and a set of momenta $q_1,q_2,\cdots$, and $|{\rm FS}\rangle$ represents the Fermi sea.
From the parity and symmetry, it is easy to find $\Omega^{iij}_{p,q}=-\Omega^{iij}_{-p,q}$ and $\Omega_{p,q}^{iij}=\Omega_{p,-q}^{iij}$ ($i,j=a,b$ and $i\neq j$) for Cooper {trimer} phases.
In addition, here $\displaystyle\sum_{p_1,p_2,\cdots}'$ is adopted to denote the momentum summation restricted by the Fermi surface for convenience. 
The step functions associated with the Fermi surface will be recovered when we numerically evaluate the momentum summation.
By minimizing the in-medium two- and three-body energies based on the variational principle, the variational parameter $\Omega^{mn\cdots}_{q_1,q_2,\cdots}$ will be determined correspondingly.
{In the present work, we assume equal particle densities for the two components, such that $k_{{\rm F},a}=k_{{\rm F}, b}\equiv k_{\rm F}$. 
As a consequence, Pauli blocking is implemented using a common Fermi momentum in Eqs.~\eqref{eq8}--\eqref{eq12}, while the chemical potentials satisfy $\mu_i=k_{\rm F}^2/(2m_i)$ and differ due to the mass imbalance.}

Calculating the expectation values of Hamiltonian associated with each trial wave function $\langle\Psi_{mn\cdots}|H|\Psi_{mn\cdots}\rangle$, and then applying the variational principle, we arrive that
\begin{align}
    \frac{\delta}{\delta {\Omega^{mn\cdots}_{q_1,{q_2},\cdots}}^*}\langle\Psi_{mn\cdots}|(H-E^{mn\cdots}_\alpha)|\Psi_{mn\cdots}\rangle=0,
\end{align}
where $E_\alpha^{mn\cdots}$ ($\alpha=2,3$) is the {lowest-lying} energy of a Cooper cluster with $mn\cdots$ configuration.
As a result,the in-medium variational equations read
\begin{align}
\label{eq:e2s}
1+U_s\sum_{q}'\frac{1}{\xi_{q,a}+\xi_{q,b}-E_{2,s}}=0,
\end{align}
for the $s$-wave pairing~\cite{Ohashi2020Prog.Part.andNucl.Phys.111.103739},
\begin{align}
\label{eq:e2p}
    1+U_i\sum_{p}'\frac{p^2}{\xi_{p,i}+\xi_{-p,i}-E_{2,p}^{ii}}=0,
\end{align}
for $p$-wave pairing ($ii$ configuration)~\cite{Guo2022Phys.Rev.A106.043310},
\begin{align}
\label{eq:e3aab}
   & (\xi_{p+q/2,a}+\xi_{-p+q/2,a}+\xi_{-q,b}-E_3^{aab})
    \Omega^{aab}_{p,q}+p\Gamma^{aab}_p(q)\nonumber\\
    &+\Gamma^{aab}_s(p-q/2)-\Gamma^{aab}_s(-p-q/2)=0,
\end{align}
for {Cooper trimer} ($aab$ configuration)~\cite{Guo2023Phys.Rev.B107.024511}, and
\begin{align}
\label{eq:e3abb}
   & (\xi_{-q,a}+\xi_{p+q/2,b}+\xi_{-p+q/2,b}-E_3^{abb})
    \Omega^{abb}_{p,q}+p\Gamma^{abb}_p(q)\nonumber\\
    &+\Gamma^{abb}_s(p-q/2)-\Gamma^{abb}_s(-p-q/2)=0,
\end{align}
for {Cooper trimer} ($abb$ configuration), respectively.
To simplify the ensuing expressions, in the above variational equations we introduce
\begin{align}
    \Gamma_p^{iij}(q)=U_i\sum_{p'}'p'\Omega^{iij}_{p',q},\quad \Gamma^{iij}_s(k)=U_s\sum_{q'}'\Omega^{iij}_{k+q'/2,q'},
\end{align}
with $i,j=a,b$ and $i\neq j$.
We also figure out that
\begin{align}\label{eq19}
    \Gamma^{iij}_p(q)=\Gamma^{iij}_{p}(-q),
    \quad
    \Gamma^{iij}_s(k)=-\Gamma^{iij}_s(-k).
\end{align}
Correspondingly, the in-medium three-body equations for $\Gamma^{iij}_p(q)$ and $\Gamma^{iij}_s(k)$ reads
\begin{align}\label{gammap}
    &\Gamma^{iij}_p(q)
    \left[
    1+U_i\sum_{p}'
    \frac{
    p^2}{\xi_{p+q/2,i}+\xi_{-p+q/2,i}+\xi_{-q,j}-E_3^{iij}}
    \right]\cr
    =\,&-U_i\sum_{p}'p\frac{\Gamma^{iij}_s(p-q/2)-\Gamma^{iij}_s(-p-q/2)
    }{\xi_{p+q/2,i}+\xi_{-p+q/2,i}+\xi_{-q,j}-E_3^{iij}},
\end{align}
and 
\begin{align}\label{gammas}
    &\Gamma^{iij}_{s}(k)
    \left[   1+U_s\sum_{q}'\frac{1}{\xi_{k+q,i}+\xi_{-k,i}+\xi_{-q,j}-E_3^{iij}}
    \right]\cr
    =\,&-U_s\sum_{q}'\frac{(k+q/2)\Gamma^{iij}_p(q)-\Gamma^{iij}_s(-k-q)
    }{\xi_{k+q,i}+\xi_{-k,i}+\xi_{-q,j}-E_3^{iij}},
\end{align}
respectively.

Before discussing the numerical demonstrations in detail, we note that present variational approach is able to describe the BCS-BEC crossover and its three-body counterpart qualitatively~\cite{Tajima2021PhysRevA.104.023319}.
The variational wave functions for calculating the energies of Cooper clusters can well depict both the Cooper clustering in the weak-coupling BCS limit and the tightly bound state formation in the strong-coupling BEC limit in a unified manner.
{We also emphasize that the present approach is a variational analysis for the generalized Cooper problem performed on top of an inert Fermi sea. 
Therefore, the obtained energies and phase diagrams should be understood as a hierarchy of lowest-energy cluster instabilities within the adopted variational subspace, rather than the fully self-consistent thermodynamic many-body ground state.}

\section{Results and discussion}\label{sec:III}

{
In this section, we are going to present the numerical implementation of analytic derivations from Eqs.~\eqref{eq:e2s}--\eqref{gammas}.
All momentum summations are evaluated on uniform grids up to a finite cutoff $\Lambda$, with a grid spacing $\mathrm{d}k = 0.01\,k_{\rm F}$, which is sufficient for numerical convergence.
The coupled equations for the auxiliary amplitudes $\Gamma^{iij}_p(q)$ and $\Gamma^{iij}_s(k)$ are solved after discretization as a linear eigenvalue problem, whose lowest eigenvalue determines the cluster energy.
The symmetry relations stated in Eq.~\eqref{eq19} are explicitly enforced in the numerical implementation to reduce the independent degrees of freedom and improve numerical stability.
Convergence is checked by refining the momentum grids, and by independently varying the cutoff $\Lambda$ while keeping the physical scattering parameters fixed.}

\subsection{In-vacuum case}\label{sec:IIIA}

In the absence of the Fermi sea, namely, by taking the reference state as the vacuum $|0\rangle$ instead of the Fermi sea $|\textrm{FS}\rangle$, we first investigate the in-vacuum two- and three-body clustering in present mass-imbalanced two-component Fermi gas.

\begin{figure}
  \includegraphics[width=1.0\linewidth]{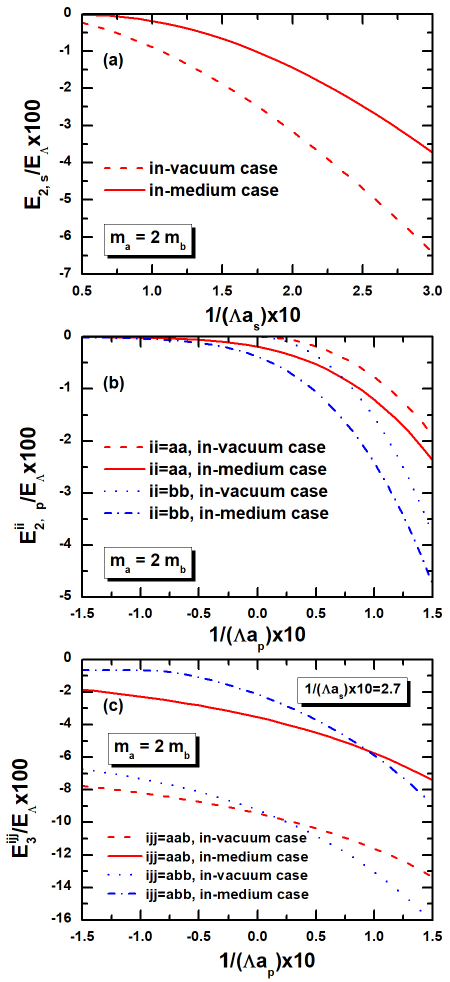}
  \caption{
  {(a)~In-vacuum and in-medium $s$-wave pairing energies as functions of the $s$-wave scattering length.
(b)~In-vacuum and in-medium $p$-wave pairing energies with different configurations as functions of the $p$-wave scattering length.
(c)~In-vacuum and in-medium three-body energies with different configurations as functions of the $p$-wave scattering length at a fixed $s$-wave interaction strength $1/(\Lambda a_s)\times 10 = 2.7$.
For the in-vacuum case, all energies are measured relative to the vacuum threshold.
For the in-medium case, energies are measured relative to the energy of the filled Fermi sea of noninteracting fermions.
A more strongly bound state corresponds to a more negative energy (or equivalently a larger $|E|$).
In all panels, the mass ratio is fixed to $m_a/m_b=2$, and the momentum cutoff is $\Lambda/k_F=10$.
The reference energy scale is $E_\Lambda=\Lambda^2/(2m_r)$.
}
  }\label{fig:4}
\end{figure}

In Fig.~\ref{fig:4}, by taking the mass ratio $m_a = 2 m_b$ as a typical example, we plot in-vacuum clustering energies as functions of scattering lengths.
As for the in-vacuum two-body energies shown in Figs.~\ref{fig:4}(a) and (b), the $s$($p$)-wave pairing one monotonically decreases with the increase of $s$($p$)-wave interaction strengths.
From Fig.~\ref{fig:4}(b), it is seen that the in-vacuum two-body pairing with $aa$ configuration is always suppressed by the one with $bb$ configuration.
It is due to the mass difference, where the pairing consisting of lighter atoms is more tightly bound at a fixed interaction strength.
Such a tendency can be qualified from Eq.~\eqref{eq:e2p}.
One has
\begin{align}
  \frac{E_{2,p}^{ii}}{E_\Lambda}
  \propto \frac{m_r}{m_i}
\end{align}
in the reference scale of $E_\Lambda=\Lambda^2/(2m_r)$, where $\Lambda$ denotes the momentum cutoff adopted in the practical calculations.
It naturally indicates $p$-wave $bb$ pairing is more tightly bound than the $aa$ configuration. 

In Fig.~\ref{fig:4}(c), by fixing $1/(\Lambda a_s)\times10=2.7$, the in-vacuum three-body energies become more negative (i.e., the binding increases) with stronger $p$-wave interaction.
Among the in-vacuum three-body clusters, the $aab$ configuration is dominant in the regime $1/(\Lambda a_p)\times10\lesssim0.25$, while the $abb$ one becomes more stable when $1/(\Lambda a_p)\times10$ exceeds $0.25$.
In other words, there is a competition between two different configurations of in-vacuum three-body clusters, which also results from the mass difference.

\begin{figure}
  \includegraphics[width=1.0\linewidth]{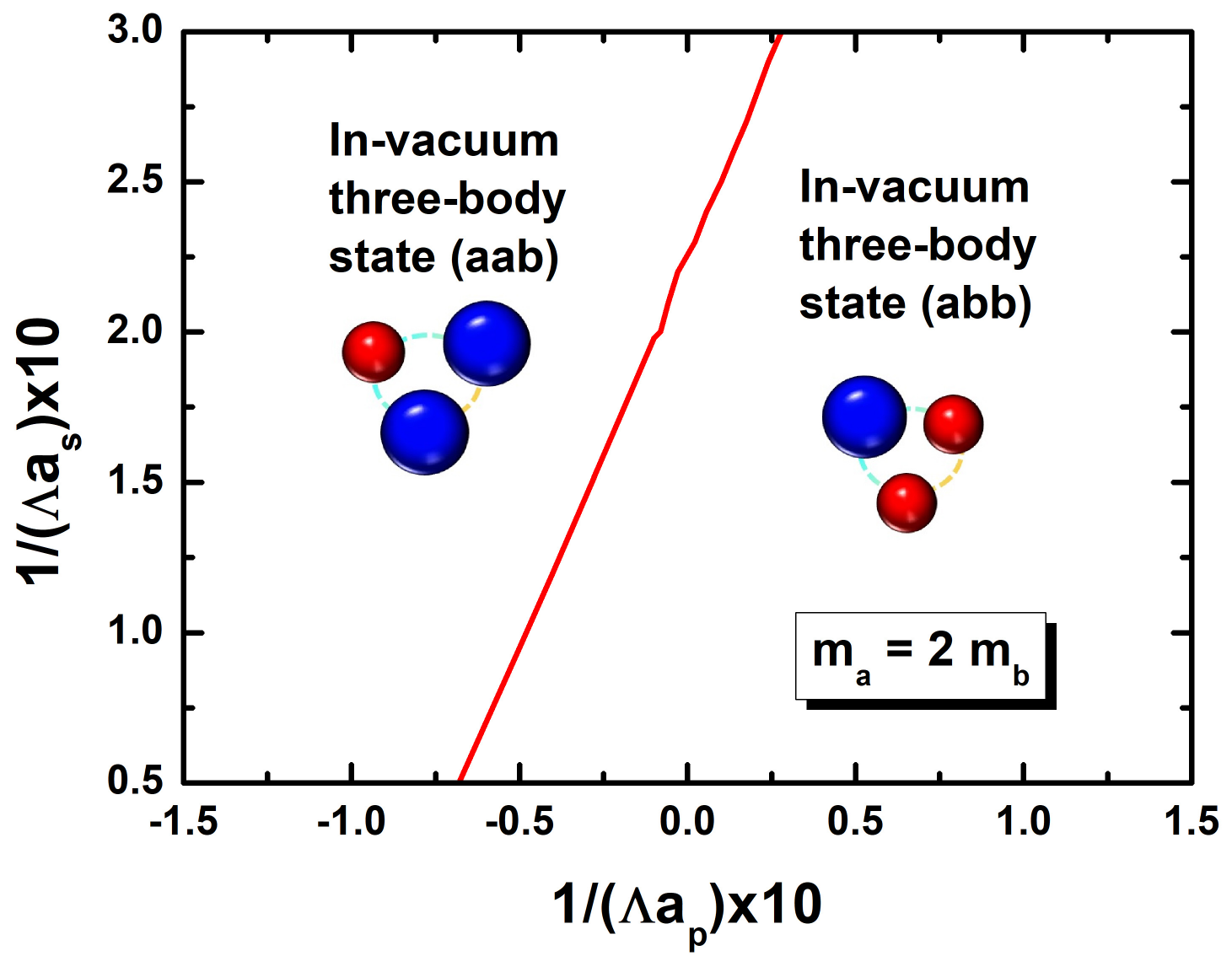}
  \caption{
  {In-vacuum phase diagram in the plane of $1/(\Lambda a_s)$ and $1/(\Lambda a_p)$.
The phase shown corresponds to the most deeply bound cluster, determined by comparing the in-vacuum two- and three-body energies.
All energies are measured relative to the vacuum threshold, and a more strongly bound state corresponds to a more negative energy.
Within the explored parameter window, the three-body states are always more deeply bound than the two-body ones.
The mass ratio is fixed to $m_a/m_b=2$, and the momentum cutoff is $\Lambda/k_F=10$.
}
    }\label{fig:invac}
\end{figure}

We numerically map out the in-vacuum phase diagram in the plane of $1/(\Lambda a_s)$ and $1/(\Lambda a_p)$ as shown in Fig.~\ref{fig:invac}.
The in-vacuum phase diagram gives the lowest-lying phase among all the five configurations of two- or three-body clusters discussed in this work.
The mass ratio is taken as $m_a=2m_b$.
{Within the explored ranges of $a_s$ and $a_p$ shown in Fig.~\ref{fig:invac}, the in-vacuum three-body states are always more deeply bound than the two-body ones.
Here the binding energy is defined as the energy measured relative to the vacuum threshold, so that a more strongly bound state corresponds to a more negative energy (or a larger $|E|$).}
{Physically, the predominance of three-body binding in vacuum can be understood as a cooperative effect of the coexistent $s$- and $p$-wave attractions.
While the $s$-wave interaction favors interspecies pairing and the $p$-wave interaction provides attraction between identical fermions, a three-body configuration can benefit from both channels simultaneously.
This additional binding mechanism is absent in isolated two-body states, leading to a deeper trimer binding within the explored parameter regime.}
Moreover, at a fixed $s$-wave interaction, with the increase of $p$-wave interaction, the lowest-lying phase changes from the three-body states with $aab$ configuration to those with $abb$ configuration.
In other words, the in-vacuum three-body states with $abb$ configuration become dominant in the regime with stronger $p$-wave interaction.
It is because the increase of $p$-wave attraction induces larger enhancement for $bb$ pairings with smaller mass compared to $aa$ ones.

\subsection{In-medium case}\label{sec:IIIB}

\begin{figure}
  \includegraphics[width=1.0\linewidth]{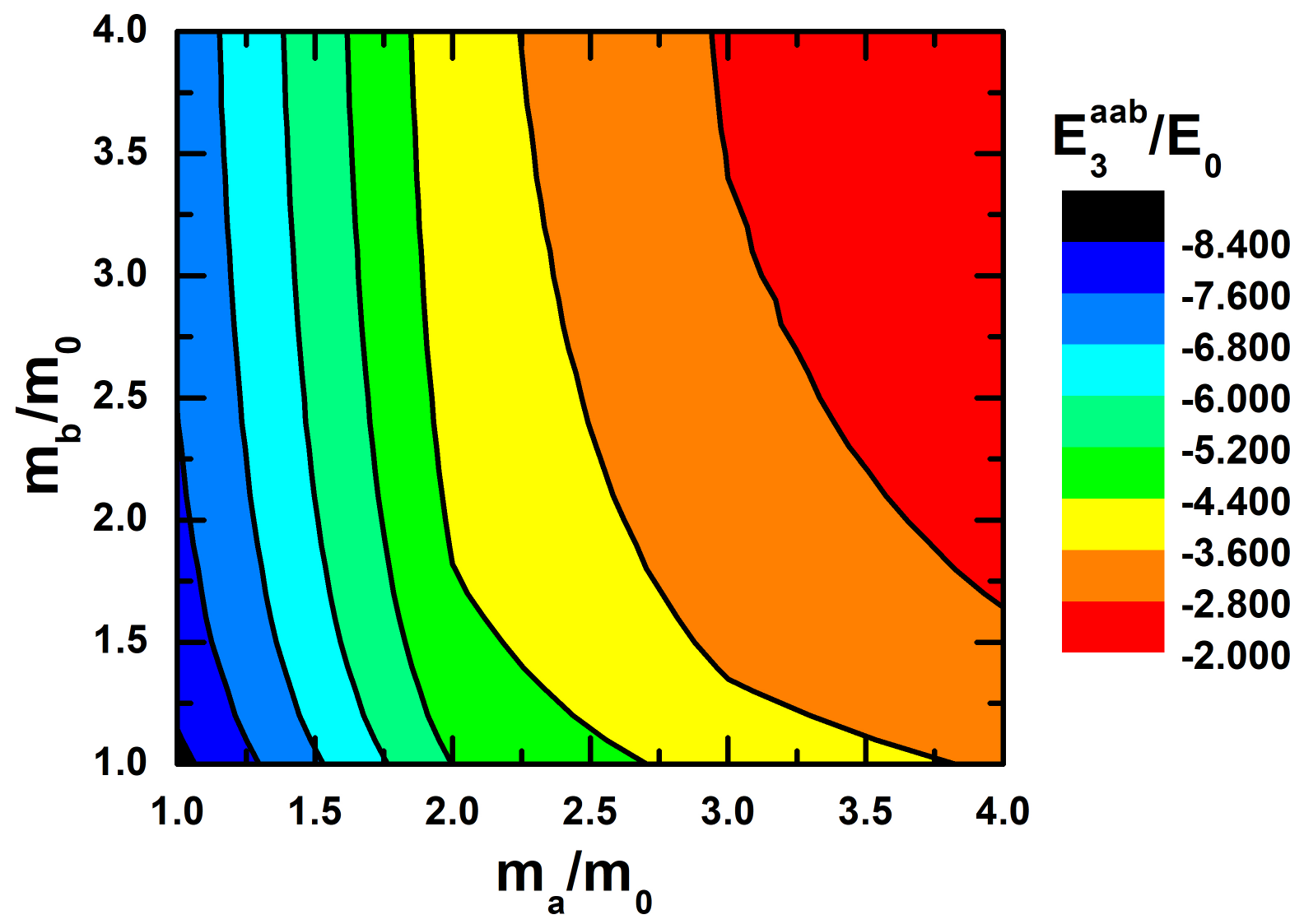}
  \caption{{Contour plot of the in-medium three-body energy $E^{aab}_3/E_0$ as a function of the masses $m_a/m_0$ and $m_b/m_0$.
Here $E_0=k_F^2/(2m_0)$ is the reference energy scale.
The energy is measured relative to the filled Fermi sea of noninteracting fermions, and a more negative value indicates a more strongly bound in-medium trimer.
The interaction strengths are fixed to $1/(k_F a_s)=1/(k_F a_p)=1.5$, and the momentum cutoff is $\Lambda/k_F=10$.}}\label{fig:1}
\end{figure}

To see effects from the mass imbalance to the in-medium three-body energy, in Fig.~\ref{fig:1}, we plot the contour plot of in-medium three-body energy $E_3^{aab}/E_0$ in a plane of mass $m_a/m_0$ and $m_b/m_0$.
For comparison, the mass $m_0$ and energy $E_0=k_{\rm F}^2/(2m_0)$ are taken as the reference scales.
The interaction strengths are fixed at $1/(k_{\rm F}a_s)=1/(k_{\rm F}a_p)=1.5$.
The momentum cutoff $\Lambda$ is taken as $10k_{\rm F}$.
In the present two-component Fermi gas, as for one species of atom with a fixed mass, if we gradually enlarge the mass of another one, the absolute value of $E_3^{aab}$ monotonically decreases.
Such a qualitative behavior is a result of fixed interaction strengths.
Moreover, it is also seen that a smaller reduced mass $m_r=m_am_b/(m_a+m_b)$ gives {a more negative} $E_3^{aab}$ in Fig.~\ref{fig:1}. 

\begin{figure}
  \includegraphics[width=1.0\linewidth]{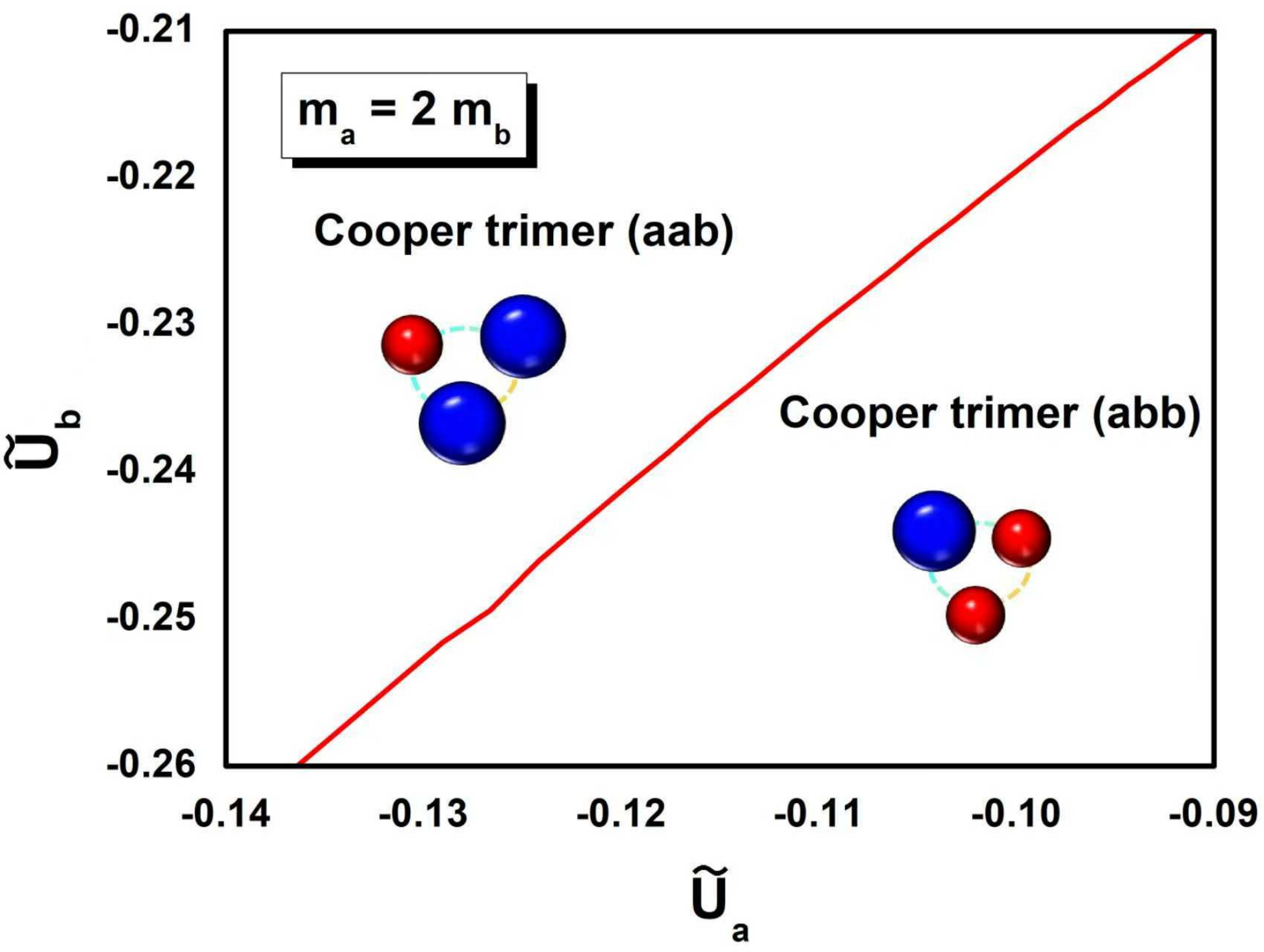}
  \caption{Lower-lying excited state of Cooper {trimer} phases with different configurations in a plane of two-body $p$-wave interaction strengths.
   The mass ratio is taken as $m_a=2m_b$, and the momentum cutoffs are taken as $\Lambda/k_{\rm F}=10$.
  }\label{fig:2}
\end{figure}

In order to see the competition between the in-medium three-body correlations with different configurations in the present system, by fixing the mass ratio as $m_a=2m_b$, we plot the lower-lying excited state of Cooper {trimer} phases with different configurations in a plane of two-body $p$-wave interaction strengths in Fig.~\ref{fig:2}.
Here the dimensionless $p$-wave coupling strengths are defined as $\tilde{U}_i=2m_rk_{\rm F}U_i$.
In this paper, we consider the attractive two-body $p$-wave interactions, which indicates that a larger absolute value of $\tilde{U}_i$ represents a stronger attraction.
At a fixed $p$-wave interaction strength $\tilde{U}_a$ among atoms $a$, when the $p$-wave interaction strength $\tilde{U}_b$ among atoms $b$ increases to a certain value, the dominant (lower-lying) Cooper {trimer} state changes from the $aab$ configuration to the $abb$ one.
The similar behavior holds for a fixed $p$-wave interaction strength $\tilde{U}_b$ among atoms $b$.
The competition between two different configurations of Cooper {trimers} is similar to the in-vacuum case shown in Fig.~\ref{fig:invac}, which results from the mass difference.

\begin{figure}
  \includegraphics[width=1.0\linewidth]{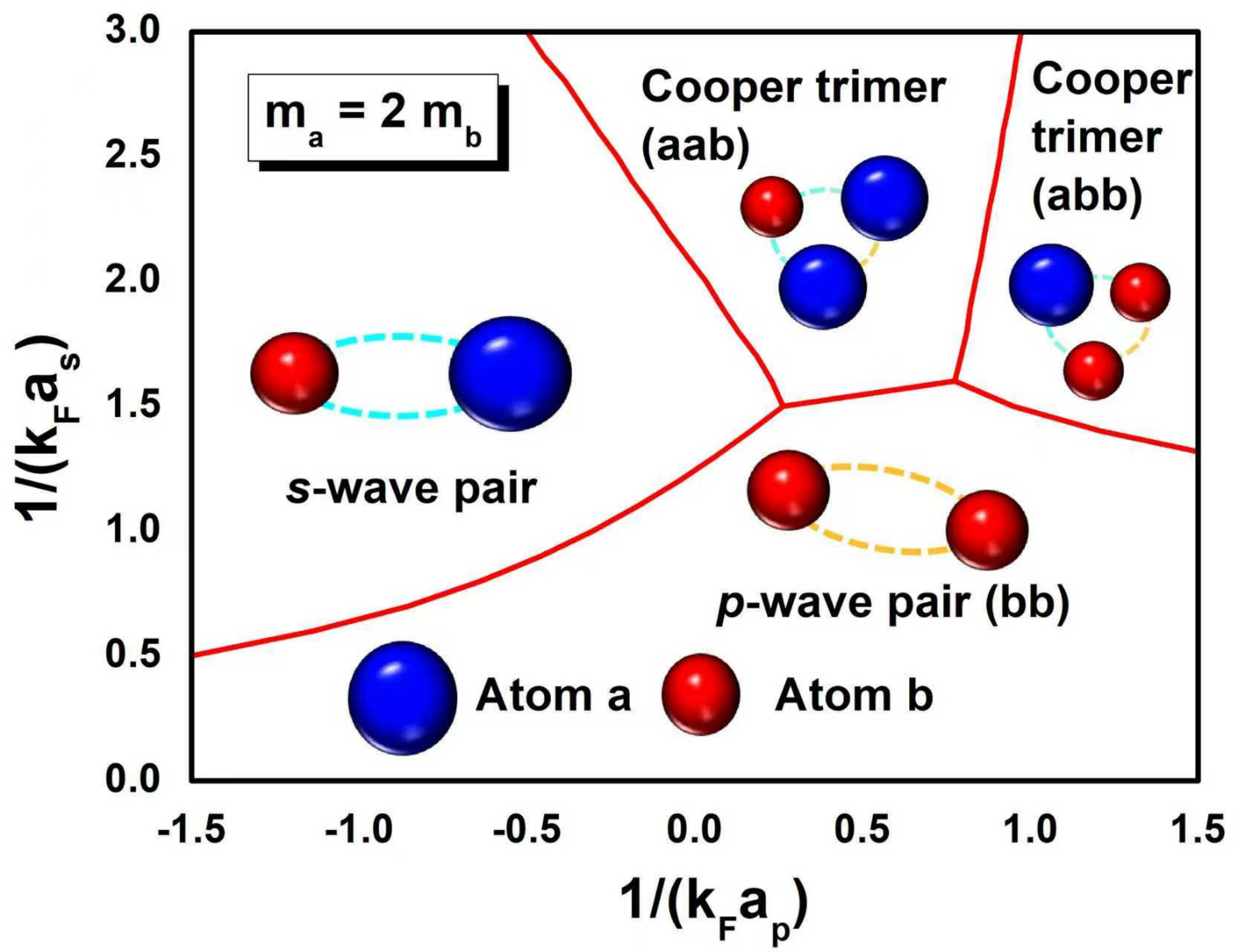}
  \caption{
  {Phase diagram of $s$-wave pairing, $p$-wave pairing, and Cooper trimer states in the plane of $1/(k_F a_s)$ and $1/(k_F a_p)$.
The phase boundaries are determined by comparing the in-medium cluster energies, with the dominant instability corresponding to the most negative energy.
All energies are measured relative to the energy of the filled Fermi sea of noninteracting fermions.
The mass ratio is fixed to $m_a/m_b=2$, and the momentum cutoff is $\Lambda/k_F=10$.}}\label{fig:3}
\end{figure}

At last, by taking the mass ratio as $m_a=2m_b$,  {we summarize the resulting hierarchy of dominant instabilities of $s$- and $p$-wave Cooper pairing states, and the Cooper {trimer} states with different configurations in Fig.~\ref{fig:3}.}
The boundary lines between different kinds of Cooper clustering phases are determined in such a way that the one between {Cooper trimer} state ($iij$ configuration) and $s$-wave pairing, that between {Cooper trimer} state ($iij$ configuration) and $p$-wave pairing ($ii$ configuration), and that between $s$- and $p$-wave pairing ($ii$ configuration) are given by $E_3^{iij}=E_{2,s}$, $E_3^{iij}=E_{2,p}^{ii}$, and $E_{2,s}=E_{2,p}^{ii}$, respectively.
The range of scattering parameters in Fig.~\ref{fig:3} [$0.5\leq 1/(k_{\rm F}a_s)\leq 3$ and $-1.5\leq 1/(k_{\rm F}a_p)\leq 1.5$] are adopted as same one in Ref.~\cite{Guo2023Phys.Rev.B107.024511}, which are also similar values were realized in a recent experimental work~\cite{Jackson2023Phys.Rev.X13.021013}.
Such a phase diagram captures interesting features associated with competing $s$- and $p$-wave pairings and moreover {trimer} formation.
In order to further investigate the underlying physics, we also plot in-medium clustering energies as functions of scattering lengths in Fig.~\ref{fig:4}.
From Fig.~\ref{fig:4}, it is shown that the presence of the Fermi sea reshapes the formation of pairing and three-body states.
Moreover, by comparing Fig.~\ref{fig:4}(a) [or (b)] and (c), the presence of Fermi surface affects three-body energies much more than pairing ones.
Consequently, it is seen that when $s$- or $p$-wave interaction strength is comparably weak, the pairing phases survive, which is different to the in-vacuum case.
As for the pairing states, the $p$-wave pair ($aa$ configuration) phase is always suppressed due to the mismatch of Fermi energies between atoms $a$ and $b$, where the Cooper instability among heavier atoms becomes weaker.
As for the {trimer} states, the Cooper {trimer} ($aab$ configuration) phase survives due to the coexistence of both $s$- and $p$-wave interactions, where the effect from the mismatch of Fermi energies is partially offset by the inclusion of $s$-wave attraction.
In the meanwhile, similar to the in-vacuum case shown in Fig.~\ref{fig:invac}, there is a competition between Cooper {trimer} ($aab$ configuration) phase and Cooper {trimer} ($abb$ configuration) phase.
On the one hand, we have numerically checked that the qualitative behaviors of the boundary line between $s$-wave pair phase and Cooper {trimer} ($aab$ configuration) phase, and that between Cooper {trimer} ($aab$ configuration) phase and Cooper {trimer} ($abb$ configuration) phase are unchanged even up to a extremely large $s$-wave interaction strength [such as $1/(k_{\rm F}a_s)=10.0$].
It indicates that the competition between different configurations of Cooper {trimer} phases may also result from the difference of scatterings between $s$-wave dimer ($ab$ configuration) and each component of atoms.
Meanwhile, at a fixed $s$-wave interaction strength, $bb$ configuration with a smaller mass also becomes more tightly bound than the $aa$ configuration as the  $p$-wave scattering length increases due to the mismatch of Fermi energy between atoms $a$ and $b$.
In other words, the Cooper {trimer} ($abb$ configuration) is expected to be more stable than the Cooper {trimer} ($aab$ configuration) in the stronger $p$-wave interaction regime.

\begin{table}[tb]
  \centering
  \caption{
   { Cutoff dependence of boundary point $1/(k_{\rm F}a_p)$ of $s$-wave pairing and Cooper trimer state ($aab$ configuration).}}
  \label{tab:1}
  \begin{ruledtabular}
    \begin{tabular}{cccc}
      $1/(k_{\rm F}a_s)$  & $\Lambda/k_{\rm F}=8$ & $\Lambda/k_{\rm F}=10$ & $\Lambda/k_{\rm F}=12$\\
      \hline 
      $2.8$ & $0.021$ &$-0.386$ & $-0.522$ \\
      $2.9$ & $-0.001$ &$-0.450$ & $-0.666$ \\
      $3.0$ & $-0.175$ &$-0.500$ & $-0.850$ \\
    \end{tabular}
  \end{ruledtabular}
\end{table}

{We have numerically checked the cutoff dependence, and the physical two-body scattering parameters $a_s$ and $a_p$ are kept fixed when varying the momentum cutoff $\Lambda$.
As shown in Table~\ref{tab:1}, by scanning several boundary points of $s$-wave pairing and Cooper trimer state ($aab$ configuration), we find that increasing $\Lambda$ enlarges the parameter region where the Cooper trimer instability dominates.
While the quantitative locations of the phase boundaries exhibit moderate cutoff-dependent shifts, the qualitative structure of the phase diagram remains robust, including the hierarchy between pairing and Cooper trimer instabilities and the competition between different Cooper trimer configurations.
The momentum cutoff is introduced as an ultraviolet regularization for the zero-range interactions. 
Even though the same cutoff corresponds to different single-particle energy scales for the two species in the mass-imbalanced case, the low-energy physics discussed here is governed by the relative momentum and remains unaffected.}
Such a cutoff dependence is also similar to that in the previous work~\cite{Guo2023Phys.Rev.B107.024511}.
While the specific value of momentum cutoff in the practical calculations is needed to compare our results with the experiments, our phase diagram would be useful to understand the qualitative features of hybridized $s$- and $p$-wave interacting systems with mass imbalance and different configurations of clustering.

{
We note that effects beyond the present variational treatment, such as medium polarization, self-energy dressing, and finite lifetime of Cooper clusters, are not included. 
These effects may quantitatively renormalize the cluster energies and phase boundaries. 
Nevertheless, as long as the system is in the regime where pairing or clustering instabilities emerge continuously from the Fermi surface, the relative ordering of instabilities is expected to remain qualitatively reliable.
Furthermore, the possible emergence of four-body bound states, such as a tetramer in $aabb$ configuration, cannot be excluded in principle, especially in the presence of both $a$-$a$ and $b$-$b$ odd-parity interactions. 
The explicit treatment of such four-body correlations is beyond the scope of the present work and is left for future studies.}

{Before drawing the conclusion, it is noted that although most results are presented for a representative mass ratio $m_a/m_b=2$, we have verified that the qualitative features of the pairing--trimer competition and the structure of the phase diagram persist for other mass ratios within a reasonable range.
In particular, varying the mass ratio mainly leads to quantitative shifts of phase boundaries, while the hierarchy between pairing and Cooper trimer formation remains unchanged.}

\section{Summary and perspectives}\label{sec:IV}

As a step further of previous work~\cite{Guo2023Phys.Rev.B107.024511}, in this paper we have investigated the mass imbalance and different kinds of clustering configurations in the system with the coexistence $s$- and $p$-wave interactions.
We have investigated both the in-vacuum and in-medium two- and three-body correlations in the mass-imbalanced two-component Fermi gas.
The solutions of in total five kinds of stable in-medium two- and three-body cluster states have been found in this system by solving the variational equations.
By taking a certain mass ratio between two components of fermions, we have {mapped out} a {lowest-energy cluster-instability diagram} consisting of $s$- and $p$-wave pairing states, and the Cooper {trimer} states with different configurations in a plane of $s$- and $p$-wave scattering lengths.

With the inclusion of mass imbalance and different clustering configurations, our work would be useful for the deeper understanding and better understanding of the nontrivial states arising in the systems with the coexistence of both even-parity ($s$-wave) and odd-parity ($p$-wave) interactions.
Present results would also be useful for the further investigation of unconventional superfluids and superconductors.
For the future perspectives, it is also interesting to see how similar in-medium correlations may arise in other systems, such as condensed-matter or nuclear ones.
For instance, our present model setup may be applied to the investigations of hypernuclei with the lightest hyperon $\Lambda$~\cite{Danysz1953}.

\begin{acknowledgments}
The author thanks Hiroyuki Tajima for useful discussions.
Y.G. was supported by RIKEN Special Postdoctoral Researchers Program.
\end{acknowledgments}

%\bibliography{GYX}

\begin{thebibliography}{47}%
\makeatletter
\providecommand \@ifxundefined [1]{%
 \@ifx{#1\undefined}
}%
\providecommand \@ifnum [1]{%
 \ifnum #1\expandafter \@firstoftwo
 \else \expandafter \@secondoftwo
 \fi
}%
\providecommand \@ifx [1]{%
 \ifx #1\expandafter \@firstoftwo
 \else \expandafter \@secondoftwo
 \fi
}%
\providecommand \natexlab [1]{#1}%
\providecommand \enquote  [1]{``#1''}%
\providecommand \bibnamefont  [1]{#1}%
\providecommand \bibfnamefont [1]{#1}%
\providecommand \citenamefont [1]{#1}%
%\providecommand \href@noop [0]{\@secondoftwo}%
%\providecommand \href [0]{\begingroup \@sanitize@url \@href}%
\providecommand \@href[1]{\@@startlink{#1}\@@href}%
\providecommand \@@href[1]{\endgroup#1\@@endlink}%
\providecommand \@sanitize@url [0]{\catcode `\\12\catcode `\$12\catcode
  `\&12\catcode `\#12\catcode `\^12\catcode `\_12\catcode `\%12\relax}%
\providecommand \@@startlink[1]{}%
\providecommand \@@endlink[0]{}%
%\providecommand \url  [0]{\begingroup\@sanitize@url \@url }%
\providecommand \@url [1]{\endgroup\@href {#1}{\urlprefix }}%
\providecommand \urlprefix  [0]{URL }%
\providecommand \doibase [0]{https://doi.org/}%
\providecommand \selectlanguage [0]{\@gobble}%
\providecommand \bibinfo  [0]{\@secondoftwo}%
\providecommand \bibfield  [0]{\@secondoftwo}%
\providecommand \translation [1]{[#1]}%
\providecommand \BibitemOpen [0]{}%
\providecommand \bibitemStop [0]{}%
\providecommand \bibitemNoStop [0]{.\EOS\space}%
\providecommand \EOS [0]{\spacefactor3000\relax}%
\providecommand \BibitemShut  [1]{\csname bibitem#1\endcsname}%
\let\auto@bib@innerbib\@empty
%</preamble>
\bibitem [{\citenamefont {Yasui}\ \emph {et~al.}(2020)\citenamefont {Yasui},
  \citenamefont {Inotani},\ and\ \citenamefont
  {Nitta}}]{Yasui2020PhysRevC.101.055806}%
  \BibitemOpen
  \bibfield  {author} {\bibinfo {author} {\bibfnamefont {S.}~\bibnamefont
  {Yasui}}, \bibinfo {author} {\bibfnamefont {D.}~\bibnamefont {Inotani}},\
  and\ \bibinfo {author} {\bibfnamefont {M.}~\bibnamefont {Nitta}},\ }\bibfield
   {title} {\bibinfo {title} {Coexistence phase of $^{1}$S$_{0}$ and
  $^{3}$P$_{2}$ superfluids in neutron stars},\ }\href
  {https://doi.org/10.1103/PhysRevC.101.055806} {\bibfield  {journal} {\bibinfo
   {journal} {Phys. Rev. C}\ }\textbf {\bibinfo {volume} {101}},\ \bibinfo
  {pages} {055806} (\bibinfo {year} {2020})}\BibitemShut {NoStop}%
\bibitem [{\citenamefont {Marqu{\'e}s}\ and\ \citenamefont
  {Carbonell}(2021)}]{marques2021quest}%
  \BibitemOpen
  \bibfield  {author} {\bibinfo {author} {\bibfnamefont {F.~M.}\ \bibnamefont
  {Marqu{\'e}s}}\ and\ \bibinfo {author} {\bibfnamefont {J.}~\bibnamefont
  {Carbonell}},\ }\bibfield  {title} {\bibinfo {title} {The quest for light
  multineutron systems},\ } {\bibfield  {journal} {\bibinfo
  {journal} {The European Physical Journal A}\ }\textbf {\bibinfo {volume}
  {57}},\ \bibinfo {pages} {1} (\bibinfo {year} {2021})}\BibitemShut {NoStop}%
\bibitem [{\citenamefont {Duer}\ \emph {et~al.}(2022)\citenamefont {Duer},
  \citenamefont {Aumann}, \citenamefont {Gernh{\"a}user}, \citenamefont
  {Panin}, \citenamefont {Paschalis}, \citenamefont {Rossi}, \citenamefont
  {Achouri}, \citenamefont {Ahn}, \citenamefont {Baba}, \citenamefont
  {Bertulani} \emph {et~al.}}]{duer2022observation}%
  \BibitemOpen
  \bibfield  {author} {\bibinfo {author} {\bibfnamefont {M.}~\bibnamefont
  {Duer}}, \bibinfo {author} {\bibfnamefont {T.}~\bibnamefont {Aumann}},
  \bibinfo {author} {\bibfnamefont {R.}~\bibnamefont {Gernh{\"a}user}},
  \bibinfo {author} {\bibfnamefont {V.}~\bibnamefont {Panin}}, \bibinfo
  {author} {\bibfnamefont {S.}~\bibnamefont {Paschalis}}, \bibinfo {author}
  {\bibfnamefont {D.}~\bibnamefont {Rossi}}, \bibinfo {author} {\bibfnamefont
  {N.}~\bibnamefont {Achouri}}, \bibinfo {author} {\bibfnamefont
  {D.}~\bibnamefont {Ahn}}, \bibinfo {author} {\bibfnamefont {H.}~\bibnamefont
  {Baba}}, \bibinfo {author} {\bibfnamefont {C.}~\bibnamefont {Bertulani}},
  \emph {et~al.},\ }\bibfield  {title} {\bibinfo {title} {Observation of a
  correlated free four-neutron system},\ }{\bibfield  {journal}
  {\bibinfo  {journal} {Nature}\ }\textbf {\bibinfo {volume} {606}},\ \bibinfo
  {pages} {678} (\bibinfo {year} {2022})}\BibitemShut {NoStop}%
\bibitem [{\citenamefont {Kanasugi}\ and\ \citenamefont
  {Yanase}(2022)}]{Kanasugi2022Comm.Phys.5.1.}%
  \BibitemOpen
  \bibfield  {author} {\bibinfo {author} {\bibfnamefont {S.}~\bibnamefont
  {Kanasugi}}\ and\ \bibinfo {author} {\bibfnamefont {Y.}~\bibnamefont
  {Yanase}},\ }\bibfield  {title} {\bibinfo {title} {Anapole superconductivity
  from $\mathcal{PT}$-symmetric mixed-parity interband pairing},\ }{\bibfield  {journal} {\bibinfo  {journal} {Commun. Phys.}\ }\textbf
  {\bibinfo {volume} {5}},\ \bibinfo {pages} {1} (\bibinfo {year}
  {2022})}\BibitemShut {NoStop}%
\bibitem [{\citenamefont {Guo}\ and\ \citenamefont
  {Tajima}(2023{\natexlab{a}})}]{Guo2023Phys.Rev.B107.024511}%
  \BibitemOpen
  \bibfield  {author} {\bibinfo {author} {\bibfnamefont {Y.}~\bibnamefont
  {Guo}}\ and\ \bibinfo {author} {\bibfnamefont {H.}~\bibnamefont {Tajima}},\
  }\bibfield  {title} {\bibinfo {title} {Competition between pairing and
  tripling in one-dimensional fermions with coexistent $s$- and $p$-wave
  interactions},\ }\href {https://doi.org/10.1103/PhysRevB.107.024511}
  {\bibfield  {journal} {\bibinfo  {journal} {Phys. Rev. B}\ }\textbf {\bibinfo
  {volume} {107}},\ \bibinfo {pages} {024511} (\bibinfo {year}
  {2023}{\natexlab{a}})}\BibitemShut {NoStop}%
\bibitem [{\citenamefont {Efimov}(1970)}]{Efimov1970Phys.Lett.B33.563--564}%
  \BibitemOpen
  \bibfield  {author} {\bibinfo {author} {\bibfnamefont {V.}~\bibnamefont
  {Efimov}},\ }\bibfield  {title} {\bibinfo {title} {Energy levels arising from
  resonant two-body forces in a three-body system},\ }\href
  {https://doi.org/https://doi.org/10.1016/0370-2693(70)90349-7} {\bibfield
  {journal} {\bibinfo  {journal} {Phys. Lett. B}\ }\textbf {\bibinfo {volume}
  {33}},\ \bibinfo {pages} {563} (\bibinfo {year} {1970})}\BibitemShut
  {NoStop}%
\bibitem [{\citenamefont {Barontini}\ \emph {et~al.}(2009)\citenamefont
  {Barontini}, \citenamefont {Weber}, \citenamefont {Rabatti}, \citenamefont
  {Catani}, \citenamefont {Thalhammer}, \citenamefont {Inguscio},\ and\
  \citenamefont {Minardi}}]{Barontini2009Phys.Rev.Lett.103.043201}%
  \BibitemOpen
  \bibfield  {author} {\bibinfo {author} {\bibfnamefont {G.}~\bibnamefont
  {Barontini}}, \bibinfo {author} {\bibfnamefont {C.}~\bibnamefont {Weber}},
  \bibinfo {author} {\bibfnamefont {F.}~\bibnamefont {Rabatti}}, \bibinfo
  {author} {\bibfnamefont {J.}~\bibnamefont {Catani}}, \bibinfo {author}
  {\bibfnamefont {G.}~\bibnamefont {Thalhammer}}, \bibinfo {author}
  {\bibfnamefont {M.}~\bibnamefont {Inguscio}},\ and\ \bibinfo {author}
  {\bibfnamefont {F.}~\bibnamefont {Minardi}},\ }\bibfield  {title} {\bibinfo
  {title} {Observation of heteronuclear atomic efimov resonances},\ }\href
  {https://doi.org/10.1103/PhysRevLett.103.043201} {\bibfield  {journal}
  {\bibinfo  {journal} {Phys. Rev. Lett.}\ }\textbf {\bibinfo {volume} {103}},\
  \bibinfo {pages} {043201} (\bibinfo {year} {2009})}\BibitemShut {NoStop}%
\bibitem [{\citenamefont {Pires}\ \emph {et~al.}(2014)\citenamefont {Pires},
  \citenamefont {Ulmanis}, \citenamefont {H\"afner}, \citenamefont {Repp},
  \citenamefont {Arias}, \citenamefont {Kuhnle},\ and\ \citenamefont
  {Weidem\"uller}}]{PhysRevLett.112.250404}%
  \BibitemOpen
  \bibfield  {author} {\bibinfo {author} {\bibfnamefont {R.}~\bibnamefont
  {Pires}}, \bibinfo {author} {\bibfnamefont {J.}~\bibnamefont {Ulmanis}},
  \bibinfo {author} {\bibfnamefont {S.}~\bibnamefont {H\"afner}}, \bibinfo
  {author} {\bibfnamefont {M.}~\bibnamefont {Repp}}, \bibinfo {author}
  {\bibfnamefont {A.}~\bibnamefont {Arias}}, \bibinfo {author} {\bibfnamefont
  {E.~D.}\ \bibnamefont {Kuhnle}},\ and\ \bibinfo {author} {\bibfnamefont
  {M.}~\bibnamefont {Weidem\"uller}},\ }\bibfield  {title} {\bibinfo {title}
  {Observation of efimov resonances in a mixture with extreme mass imbalance},\
  }\href {https://doi.org/10.1103/PhysRevLett.112.250404} {\bibfield  {journal}
  {\bibinfo  {journal} {Phys. Rev. Lett.}\ }\textbf {\bibinfo {volume} {112}},\
  \bibinfo {pages} {250404} (\bibinfo {year} {2014})}\BibitemShut {NoStop}%
\bibitem [{\citenamefont {Fedorov}\ \emph {et~al.}(1994)\citenamefont
  {Fedorov}, \citenamefont {Jensen},\ and\ \citenamefont
  {Riisager}}]{PhysRevLett.73.2817}%
  \BibitemOpen
  \bibfield  {author} {\bibinfo {author} {\bibfnamefont {D.~V.}\ \bibnamefont
  {Fedorov}}, \bibinfo {author} {\bibfnamefont {A.~S.}\ \bibnamefont
  {Jensen}},\ and\ \bibinfo {author} {\bibfnamefont {K.}~\bibnamefont
  {Riisager}},\ }\bibfield  {title} {\bibinfo {title} {Efimov states in halo
  nuclei},\ }\href {https://doi.org/10.1103/PhysRevLett.73.2817} {\bibfield
  {journal} {\bibinfo  {journal} {Phys. Rev. Lett.}\ }\textbf {\bibinfo
  {volume} {73}},\ \bibinfo {pages} {2817} (\bibinfo {year}
  {1994})}\BibitemShut {NoStop}%
\bibitem [{\citenamefont {Jiao}\ \emph {et~al.}(2020)\citenamefont {Jiao},
  \citenamefont {Howard}, \citenamefont {Ran}, \citenamefont {Wang},
  \citenamefont {Rodriguez}, \citenamefont {Sigrist}, \citenamefont {Wang},
  \citenamefont {Butch},\ and\ \citenamefont
  {Madhavan}}]{Jiao2020Nature579.523--527}%
  \BibitemOpen
  \bibfield  {author} {\bibinfo {author} {\bibfnamefont {L.}~\bibnamefont
  {Jiao}}, \bibinfo {author} {\bibfnamefont {S.}~\bibnamefont {Howard}},
  \bibinfo {author} {\bibfnamefont {S.}~\bibnamefont {Ran}}, \bibinfo {author}
  {\bibfnamefont {Z.}~\bibnamefont {Wang}}, \bibinfo {author} {\bibfnamefont
  {J.~O.}\ \bibnamefont {Rodriguez}}, \bibinfo {author} {\bibfnamefont
  {M.}~\bibnamefont {Sigrist}}, \bibinfo {author} {\bibfnamefont
  {Z.}~\bibnamefont {Wang}}, \bibinfo {author} {\bibfnamefont {N.~P.}\
  \bibnamefont {Butch}},\ and\ \bibinfo {author} {\bibfnamefont
  {V.}~\bibnamefont {Madhavan}},\ }\bibfield  {title} {\bibinfo {title} {Chiral
  superconductivity in heavy-fermion metal ute$_2$},\ }\href
  {https://doi.org/10.1038/s41586-020-2122-2} {\bibfield  {journal} {\bibinfo
  {journal} {Nature}\ }\textbf {\bibinfo {volume} {579}},\ \bibinfo {pages}
  {523} (\bibinfo {year} {2020})}\BibitemShut {NoStop}%
\bibitem [{\citenamefont {Sauls}(1994)}]{Sauls1994Adv.Phys.43.113--141}%
  \BibitemOpen
  \bibfield  {author} {\bibinfo {author} {\bibfnamefont {J.}~\bibnamefont
  {Sauls}},\ }\bibfield  {title} {\bibinfo {title} {The order parameter for the
  superconducting phases of UPt$_3$},\ }\href
  {https://doi.org/10.1080/00018739400101475} {\bibfield  {journal} {\bibinfo
  {journal} {Adv. Phys.}\ }\textbf {\bibinfo {volume} {43}},\ \bibinfo {pages}
  {113} (\bibinfo {year} {1994})},\href{https://doi.org/10.1080/00018739400101475} \BibitemShut {NoStop}%
\bibitem [{\citenamefont {Jia}\ \emph {et~al.}(2022)\citenamefont {Jia},
  \citenamefont {Wang}, \citenamefont {Chiu}, \citenamefont {Song},
  \citenamefont {Yu}, \citenamefont {JÃ€ck}, \citenamefont {Lei}, \citenamefont
  {Klemenz}, \citenamefont {Cevallos}, \citenamefont {Onyszczak}, \citenamefont
  {Fishchenko}, \citenamefont {Liu}, \citenamefont {Farahi}, \citenamefont
  {Xie}, \citenamefont {Xu}, \citenamefont {Watanabe}, \citenamefont
  {Taniguchi}, \citenamefont {Bernevig}, \citenamefont {Cava}, \citenamefont
  {Schoop}, \citenamefont {Yazdani},\ and\ \citenamefont
  {Wu}}]{Jia2022Nat.Phys.18.87--93}%
  \BibitemOpen
  \bibfield  {author} {\bibinfo {author} {\bibfnamefont {Y.}~\bibnamefont
  {Jia}}, \bibinfo {author} {\bibfnamefont {P.}~\bibnamefont {Wang}}, \bibinfo
  {author} {\bibfnamefont {C.-L.}\ \bibnamefont {Chiu}}, \bibinfo {author}
  {\bibfnamefont {Z.}~\bibnamefont {Song}}, \bibinfo {author} {\bibfnamefont
  {G.}~\bibnamefont {Yu}}, \bibinfo {author} {\bibfnamefont {B.}~\bibnamefont
  {JÃ€ck}}, \bibinfo {author} {\bibfnamefont {S.}~\bibnamefont {Lei}}, \bibinfo
  {author} {\bibfnamefont {S.}~\bibnamefont {Klemenz}}, \bibinfo {author}
  {\bibfnamefont {F.~A.}\ \bibnamefont {Cevallos}}, \bibinfo {author}
  {\bibfnamefont {M.}~\bibnamefont {Onyszczak}}, \bibinfo {author}
  {\bibfnamefont {N.}~\bibnamefont {Fishchenko}}, \bibinfo {author}
  {\bibfnamefont {X.}~\bibnamefont {Liu}}, \bibinfo {author} {\bibfnamefont
  {G.}~\bibnamefont {Farahi}}, \bibinfo {author} {\bibfnamefont
  {F.}~\bibnamefont {Xie}}, \bibinfo {author} {\bibfnamefont {Y.}~\bibnamefont
  {Xu}}, \bibinfo {author} {\bibfnamefont {K.}~\bibnamefont {Watanabe}},
  \bibinfo {author} {\bibfnamefont {T.}~\bibnamefont {Taniguchi}}, \bibinfo
  {author} {\bibfnamefont {B.~A.}\ \bibnamefont {Bernevig}}, \bibinfo {author}
  {\bibfnamefont {R.~J.}\ \bibnamefont {Cava}}, \bibinfo {author}
  {\bibfnamefont {L.~M.}\ \bibnamefont {Schoop}}, \bibinfo {author}
  {\bibfnamefont {A.}~\bibnamefont {Yazdani}},\ and\ \bibinfo {author}
  {\bibfnamefont {S.}~\bibnamefont {Wu}},\ }\bibfield  {title} {\bibinfo
  {title} {Evidence for a monolayer excitonic insulator},\ }\href
  {https://doi.org/10.1038/s41567-021-01422-w} {\bibfield  {journal} {\bibinfo
  {journal} {Nature Phys.}\ }\textbf {\bibinfo {volume} {18}},\ \bibinfo {pages}
  {87} (\bibinfo {year} {2022})}\BibitemShut {NoStop}%
\bibitem [{\citenamefont {Mackenzie}\ and\ \citenamefont
  {Maeno}(2003)}]{Mackenzie2003Rev.Mod.Phys.75.657--712}%
  \BibitemOpen
  \bibfield  {author} {\bibinfo {author} {\bibfnamefont {A.~P.}\ \bibnamefont
  {Mackenzie}}\ and\ \bibinfo {author} {\bibfnamefont {Y.}~\bibnamefont
  {Maeno}},\ }\bibfield  {title} {\bibinfo {title} {The superconductivity of
  Sr$_2$RuO$_{4}$ and the physics of spin-triplet pairing},\ }\href
  {https://doi.org/10.1103/RevModPhys.75.657} {\bibfield  {journal} {\bibinfo
  {journal} {Rev. Mod. Phys.}\ }\textbf {\bibinfo {volume} {75}},\ \bibinfo
  {pages} {657} (\bibinfo {year} {2003})}\BibitemShut {NoStop}%
\bibitem [{\citenamefont {Kinjo}\ \emph {et~al.}(2022)\citenamefont {Kinjo},
  \citenamefont {Manago}, \citenamefont {Kitagawa}, \citenamefont {Mao},
  \citenamefont {Yonezawa}, \citenamefont {Maeno},\ and\ \citenamefont
  {Ishida}}]{Kinjo2022Science376.397--400}%
  \BibitemOpen
  \bibfield  {author} {\bibinfo {author} {\bibfnamefont {K.}~\bibnamefont
  {Kinjo}}, \bibinfo {author} {\bibfnamefont {M.}~\bibnamefont {Manago}},
  \bibinfo {author} {\bibfnamefont {S.}~\bibnamefont {Kitagawa}}, \bibinfo
  {author} {\bibfnamefont {Z.~Q.}\ \bibnamefont {Mao}}, \bibinfo {author}
  {\bibfnamefont {S.}~\bibnamefont {Yonezawa}}, \bibinfo {author}
  {\bibfnamefont {Y.}~\bibnamefont {Maeno}},\ and\ \bibinfo {author}
  {\bibfnamefont {K.}~\bibnamefont {Ishida}},\ }\bibfield  {title} {\bibinfo
  {title} {Superconducting spin smecticity evidencing the
  Fulde-Ferrell-Larkin-Ovchinnikov state in Sr$_2$RuO$_4$},\ }\href
  {https://doi.org/10.1126/science.abb0332} {\bibfield  {journal} {\bibinfo
  {journal} {Science}\ }\textbf {\bibinfo {volume} {376}},\ \bibinfo {pages}
  {397} (\bibinfo {year} {2022})}\BibitemShut {NoStop}%
\bibitem [{\citenamefont {Danysz}\ and\ \citenamefont
  {Pniewski}(1953)}]{Danysz1953}%
  \BibitemOpen
  \bibfield  {author} {\bibinfo {author} {\bibfnamefont {M.}~\bibnamefont
  {Danysz}}\ and\ \bibinfo {author} {\bibfnamefont {J.}~\bibnamefont
  {Pniewski}},\ }\bibfield  {title} {\bibinfo {title} {Delayed disintegration
  of a heavy nuclear fragment: I},\ }\href
  {https://doi.org/10.1080/14786440308520318} {\bibfield  {journal} {\bibinfo
  {journal} {The London, Edinburgh, and Dublin Philosophical Magazine and
  Journal of Science}\ }\textbf {\bibinfo {volume} {44}},\ \bibinfo {pages}
  {348} (\bibinfo {year} {1953})}\BibitemShut {NoStop}%
\bibitem [{\citenamefont {Tanida}\ \emph {et~al.}(2001)\citenamefont {Tanida},
  \citenamefont {Tamura}, \citenamefont {Abe}, \citenamefont {Akikawa},
  \citenamefont {Araki}, \citenamefont {Bhang}, \citenamefont {Endo},
  \citenamefont {Fujii}, \citenamefont {Fukuda}, \citenamefont {Hashimoto},
  \citenamefont {Imai}, \citenamefont {Hotchi}, \citenamefont {Kakiguchi},
  \citenamefont {Kim}, \citenamefont {Kim}, \citenamefont {Miyoshi},
  \citenamefont {Murakami}, \citenamefont {Nagae}, \citenamefont {Noumi},
  \citenamefont {Outa}, \citenamefont {Ozawa}, \citenamefont {Saito},
  \citenamefont {Sasao}, \citenamefont {Sato}, \citenamefont {Satoh},
  \citenamefont {Sawafta}, \citenamefont {Sekimoto}, \citenamefont {Takahashi},
  \citenamefont {Tang}, \citenamefont {Xia}, \citenamefont {Zhou},\ and\
  \citenamefont {Zhu}}]{PhysRevLett.86.1982}%
  \BibitemOpen
  \bibfield  {author} {\bibinfo {author} {\bibfnamefont {K.}~\bibnamefont
  {Tanida}}, \bibinfo {author} {\bibfnamefont {H.}~\bibnamefont {Tamura}},
  \bibinfo {author} {\bibfnamefont {D.}~\bibnamefont {Abe}}, \bibinfo {author}
  {\bibfnamefont {H.}~\bibnamefont {Akikawa}}, \bibinfo {author} {\bibfnamefont
  {K.}~\bibnamefont {Araki}}, \bibinfo {author} {\bibfnamefont
  {H.}~\bibnamefont {Bhang}}, \bibinfo {author} {\bibfnamefont
  {T.}~\bibnamefont {Endo}}, \bibinfo {author} {\bibfnamefont {Y.}~\bibnamefont
  {Fujii}}, \bibinfo {author} {\bibfnamefont {T.}~\bibnamefont {Fukuda}},
  \bibinfo {author} {\bibfnamefont {O.}~\bibnamefont {Hashimoto}}, \bibinfo
  {author} {\bibfnamefont {K.}~\bibnamefont {Imai}}, \bibinfo {author}
  {\bibfnamefont {H.}~\bibnamefont {Hotchi}}, \bibinfo {author} {\bibfnamefont
  {Y.}~\bibnamefont {Kakiguchi}}, \bibinfo {author} {\bibfnamefont {J.~H.}\
  \bibnamefont {Kim}}, \bibinfo {author} {\bibfnamefont {Y.~D.}\ \bibnamefont
  {Kim}}, \bibinfo {author} {\bibfnamefont {T.}~\bibnamefont {Miyoshi}},
  \bibinfo {author} {\bibfnamefont {T.}~\bibnamefont {Murakami}}, \bibinfo
  {author} {\bibfnamefont {T.}~\bibnamefont {Nagae}}, \bibinfo {author}
  {\bibfnamefont {H.}~\bibnamefont {Noumi}}, \bibinfo {author} {\bibfnamefont
  {H.}~\bibnamefont {Outa}}, \bibinfo {author} {\bibfnamefont {K.}~\bibnamefont
  {Ozawa}}, \bibinfo {author} {\bibfnamefont {T.}~\bibnamefont {Saito}},
  \bibinfo {author} {\bibfnamefont {J.}~\bibnamefont {Sasao}}, \bibinfo
  {author} {\bibfnamefont {Y.}~\bibnamefont {Sato}}, \bibinfo {author}
  {\bibfnamefont {S.}~\bibnamefont {Satoh}}, \bibinfo {author} {\bibfnamefont
  {R.~I.}\ \bibnamefont {Sawafta}}, \bibinfo {author} {\bibfnamefont
  {M.}~\bibnamefont {Sekimoto}}, \bibinfo {author} {\bibfnamefont
  {T.}~\bibnamefont {Takahashi}}, \bibinfo {author} {\bibfnamefont
  {L.}~\bibnamefont {Tang}}, \bibinfo {author} {\bibfnamefont {H.~H.}\
  \bibnamefont {Xia}}, \bibinfo {author} {\bibfnamefont {S.~H.}\ \bibnamefont
  {Zhou}},\ and\ \bibinfo {author} {\bibfnamefont {L.~H.}\ \bibnamefont
  {Zhu}},\ }\bibfield  {title} {\bibinfo {title} {Measurement of the
  $\mathit{B}(\mathit{E}2)$ of ${}_{\ensuremath{\Lambda}}^{7}\mathrm{Li}$ and
  shrinkage of the hypernuclear size},\ }\href
  {https://doi.org/10.1103/PhysRevLett.86.1982} {\bibfield  {journal} {\bibinfo
   {journal} {Phys. Rev. Lett.}\ }\textbf {\bibinfo {volume} {86}},\ \bibinfo
  {pages} {1982} (\bibinfo {year} {2001})}\BibitemShut {NoStop}%
\bibitem [{\citenamefont {Chin}\ \emph {et~al.}(2010)\citenamefont {Chin},
  \citenamefont {Grimm}, \citenamefont {Julienne},\ and\ \citenamefont
  {Tiesinga}}]{Chin2010Rev.Mod.Phys.82.1225--1286}%
  \BibitemOpen
  \bibfield  {author} {\bibinfo {author} {\bibfnamefont {C.}~\bibnamefont
  {Chin}}, \bibinfo {author} {\bibfnamefont {R.}~\bibnamefont {Grimm}},
  \bibinfo {author} {\bibfnamefont {P.}~\bibnamefont {Julienne}},\ and\
  \bibinfo {author} {\bibfnamefont {E.}~\bibnamefont {Tiesinga}},\ }\bibfield
  {title} {\bibinfo {title} {Feshbach resonances in ultracold gases},\ }\href
  {https://doi.org/10.1103/RevModPhys.82.1225} {\bibfield  {journal} {\bibinfo
  {journal} {Rev. Mod. Phys.}\ }\textbf {\bibinfo {volume} {82}},\ \bibinfo
  {pages} {1225} (\bibinfo {year} {2010})}\BibitemShut {NoStop}%
\bibitem [{\citenamefont {Ohashi}\ \emph
  {et~al.}(2020{\natexlab{a}})\citenamefont {Ohashi}, \citenamefont {Tajima},\
  and\ \citenamefont {{van Wyk}}}]{Ohashi2020Prog.Part.Nucl.Phys.111.103739}%
  \BibitemOpen
  \bibfield  {author} {\bibinfo {author} {\bibfnamefont {Y.}~\bibnamefont
  {Ohashi}}, \bibinfo {author} {\bibfnamefont {H.}~\bibnamefont {Tajima}},\
  and\ \bibinfo {author} {\bibfnamefont {P.}~\bibnamefont {{van Wyk}}},\
  }\bibfield  {title} {\bibinfo {title} {Bcsâbec crossover in cold atomic and
  in nuclear systems},\ }\href
  {https://doi.org/https://doi.org/10.1016/j.ppnp.2019.103739} {\bibfield
  {journal} {\bibinfo  {journal} {Prog. Part. Nucl. Phys.}\ }\textbf {\bibinfo
  {volume} {111}},\ \bibinfo {pages} {103739} (\bibinfo {year}
  {2020}{\natexlab{a}})}\BibitemShut {NoStop}%
\bibitem [{\citenamefont {Ticknor}\ \emph {et~al.}(2004)\citenamefont
  {Ticknor}, \citenamefont {Regal}, \citenamefont {Jin},\ and\ \citenamefont
  {Bohn}}]{Ticknor2004Phys.Rev.A69.042712}%
  \BibitemOpen
  \bibfield  {author} {\bibinfo {author} {\bibfnamefont {C.}~\bibnamefont
  {Ticknor}}, \bibinfo {author} {\bibfnamefont {C.~A.}\ \bibnamefont {Regal}},
  \bibinfo {author} {\bibfnamefont {D.~S.}\ \bibnamefont {Jin}},\ and\ \bibinfo
  {author} {\bibfnamefont {J.~L.}\ \bibnamefont {Bohn}},\ }\bibfield  {title}
  {\bibinfo {title} {Multiplet structure of feshbach resonances in nonzero
  partial waves},\ }\href {https://doi.org/10.1103/PhysRevA.69.042712}
  {\bibfield  {journal} {\bibinfo  {journal} {Phys. Rev. A}\ }\textbf {\bibinfo
  {volume} {69}},\ \bibinfo {pages} {042712} (\bibinfo {year}
  {2004})}\BibitemShut {NoStop}%
\bibitem [{\citenamefont {Gurarie}\ \emph {et~al.}(2005)\citenamefont
  {Gurarie}, \citenamefont {Radzihovsky},\ and\ \citenamefont
  {Andreev}}]{Gurarie2005Phys.Rev.Lett.94.230403}%
  \BibitemOpen
  \bibfield  {author} {\bibinfo {author} {\bibfnamefont {V.}~\bibnamefont
  {Gurarie}}, \bibinfo {author} {\bibfnamefont {L.}~\bibnamefont
  {Radzihovsky}},\ and\ \bibinfo {author} {\bibfnamefont {A.~V.}\ \bibnamefont
  {Andreev}},\ }\bibfield  {title} {\bibinfo {title} {Quantum phase transitions
  across a $p$-wave feshbach resonance},\ }\href
  {https://doi.org/10.1103/PhysRevLett.94.230403} {\bibfield  {journal}
  {\bibinfo  {journal} {Phys. Rev. Lett.}\ }\textbf {\bibinfo {volume} {94}},\
  \bibinfo {pages} {230403} (\bibinfo {year} {2005})}\BibitemShut {NoStop}%
\bibitem [{\citenamefont {Schunck}\ \emph {et~al.}(2005)\citenamefont
  {Schunck}, \citenamefont {Zwierlein}, \citenamefont {Stan}, \citenamefont
  {Raupach}, \citenamefont {Ketterle}, \citenamefont {Simoni}, \citenamefont
  {Tiesinga}, \citenamefont {Williams},\ and\ \citenamefont
  {Julienne}}]{Schunck2005Phys.Rev.A71.045601}%
  \BibitemOpen
  \bibfield  {author} {\bibinfo {author} {\bibfnamefont {C.~H.}\ \bibnamefont
  {Schunck}}, \bibinfo {author} {\bibfnamefont {M.~W.}\ \bibnamefont
  {Zwierlein}}, \bibinfo {author} {\bibfnamefont {C.~A.}\ \bibnamefont {Stan}},
  \bibinfo {author} {\bibfnamefont {S.~M.~F.}\ \bibnamefont {Raupach}},
  \bibinfo {author} {\bibfnamefont {W.}~\bibnamefont {Ketterle}}, \bibinfo
  {author} {\bibfnamefont {A.}~\bibnamefont {Simoni}}, \bibinfo {author}
  {\bibfnamefont {E.}~\bibnamefont {Tiesinga}}, \bibinfo {author}
  {\bibfnamefont {C.~J.}\ \bibnamefont {Williams}},\ and\ \bibinfo {author}
  {\bibfnamefont {P.~S.}\ \bibnamefont {Julienne}},\ }\bibfield  {title}
  {\bibinfo {title} {Feshbach resonances in fermionic $^{6}\mathrm{Li}$},\
  }\href {https://doi.org/10.1103/PhysRevA.71.045601} {\bibfield  {journal}
  {\bibinfo  {journal} {Phys. Rev. A}\ }\textbf {\bibinfo {volume} {71}},\
  \bibinfo {pages} {045601} (\bibinfo {year} {2005})}\BibitemShut {NoStop}%
\bibitem [{\citenamefont {Inada}\ \emph {et~al.}(2008)\citenamefont {Inada},
  \citenamefont {Horikoshi}, \citenamefont {Nakajima}, \citenamefont
  {Kuwata-Gonokami}, \citenamefont {Ueda},\ and\ \citenamefont
  {Mukaiyama}}]{Inada2008PhysRevLett.101.100401}%
  \BibitemOpen
  \bibfield  {author} {\bibinfo {author} {\bibfnamefont {Y.}~\bibnamefont
  {Inada}}, \bibinfo {author} {\bibfnamefont {M.}~\bibnamefont {Horikoshi}},
  \bibinfo {author} {\bibfnamefont {S.}~\bibnamefont {Nakajima}}, \bibinfo
  {author} {\bibfnamefont {M.}~\bibnamefont {Kuwata-Gonokami}}, \bibinfo
  {author} {\bibfnamefont {M.}~\bibnamefont {Ueda}},\ and\ \bibinfo {author}
  {\bibfnamefont {T.}~\bibnamefont {Mukaiyama}},\ }\bibfield  {title} {\bibinfo
  {title} {Collisional properties of $p$-wave feshbach molecules},\ }\href
  {https://doi.org/10.1103/PhysRevLett.101.100401} {\bibfield  {journal}
  {\bibinfo  {journal} {Phys. Rev. Lett.}\ }\textbf {\bibinfo {volume} {101}},\
  \bibinfo {pages} {100401} (\bibinfo {year} {2008})}\BibitemShut {NoStop}%
\bibitem [{\citenamefont {Nakasuji}\ \emph {et~al.}(2013)\citenamefont
  {Nakasuji}, \citenamefont {Yoshida},\ and\ \citenamefont
  {Mukaiyama}}]{Nakasuji2013PhysRevA.88.012710}%
  \BibitemOpen
  \bibfield  {author} {\bibinfo {author} {\bibfnamefont {T.}~\bibnamefont
  {Nakasuji}}, \bibinfo {author} {\bibfnamefont {J.}~\bibnamefont {Yoshida}},\
  and\ \bibinfo {author} {\bibfnamefont {T.}~\bibnamefont {Mukaiyama}},\
  }\bibfield  {title} {\bibinfo {title} {Experimental determination of $p$-wave
  scattering parameters in ultracold ${}^{6}$Li atoms},\ }\href
  {https://doi.org/10.1103/PhysRevA.88.012710} {\bibfield  {journal} {\bibinfo
  {journal} {Phys. Rev. A}\ }\textbf {\bibinfo {volume} {88}},\ \bibinfo
  {pages} {012710} (\bibinfo {year} {2013})}\BibitemShut {NoStop}%
\bibitem [{\citenamefont {Strinati}\ \emph {et~al.}(2018)\citenamefont
  {Strinati}, \citenamefont {Pieri}, \citenamefont {RÃ¶pke}, \citenamefont
  {Schuck},\ and\ \citenamefont {Urban}}]{Strinati2018Phys.Rep.738.1--76}%
  \BibitemOpen
  \bibfield  {author} {\bibinfo {author} {\bibfnamefont {G.~C.}\ \bibnamefont
  {Strinati}}, \bibinfo {author} {\bibfnamefont {P.}~\bibnamefont {Pieri}},
  \bibinfo {author} {\bibfnamefont {G.}~\bibnamefont {R\"opke}}, \bibinfo
  {author} {\bibfnamefont {P.}~\bibnamefont {Schuck}},\ and\ \bibinfo {author}
  {\bibfnamefont {M.}~\bibnamefont {Urban}},\ }\bibfield  {title} {\bibinfo
  {title} {The BCS-BEC crossover: From ultra-cold Fermi gases to nuclear
  systems},\ }\href
  {https://doi.org/https://doi.org/10.1016/j.physrep.2018.02.004} {\bibfield
  {journal} {\bibinfo  {journal} {Phys. Rep.}\ }\textbf {\bibinfo {volume}
  {738}},\ \bibinfo {pages} {1} (\bibinfo {year} {2018})}\BibitemShut {NoStop}%
\bibitem [{\citenamefont {Regal}\ \emph {et~al.}(2003)\citenamefont {Regal},
  \citenamefont {Ticknor}, \citenamefont {Bohn},\ and\ \citenamefont
  {Jin}}]{Regal2003PhysRevLett.90.053201}%
  \BibitemOpen
  \bibfield  {author} {\bibinfo {author} {\bibfnamefont {C.~A.}\ \bibnamefont
  {Regal}}, \bibinfo {author} {\bibfnamefont {C.}~\bibnamefont {Ticknor}},
  \bibinfo {author} {\bibfnamefont {J.~L.}\ \bibnamefont {Bohn}},\ and\
  \bibinfo {author} {\bibfnamefont {D.~S.}\ \bibnamefont {Jin}},\ }\bibfield
  {title} {\bibinfo {title} {Tuning $p$-wave interactions in an ultracold fermi
  gas of atoms},\ }\href {https://doi.org/10.1103/PhysRevLett.90.053201}
  {\bibfield  {journal} {\bibinfo  {journal} {Phys. Rev. Lett.}\ }\textbf
  {\bibinfo {volume} {90}},\ \bibinfo {pages} {053201} (\bibinfo {year}
  {2003})}\BibitemShut {NoStop}%
\bibitem [{\citenamefont {Zhou}\ \emph {et~al.}(2017)\citenamefont {Zhou},
  \citenamefont {Yi},\ and\ \citenamefont {Cui}}]{Zhou2017ScienceChina.60.12}%
  \BibitemOpen
  \bibfield  {author} {\bibinfo {author} {\bibfnamefont {L.}~\bibnamefont
  {Zhou}}, \bibinfo {author} {\bibfnamefont {W.}~\bibnamefont {Yi}},\ and\
  \bibinfo {author} {\bibfnamefont {X.}~\bibnamefont {Cui}},\ }\bibfield
  {title} {\bibinfo {title} {Fermion superfluid with hybridized $s$-and $p$-wave pairings},\ }{\bibfield  {journal} {\bibinfo  {journal}
  {Sci. China}\ }\textbf {\bibinfo {volume}
  {60}},\ \bibinfo {pages} {1} (\bibinfo {year} {2017})}\BibitemShut {NoStop}%
\bibitem [{\citenamefont {Niemann}\ and\ \citenamefont
  {Hammer}(2012)}]{Niemann2012Phys.Rev.A86.013628}%
  \BibitemOpen
  \bibfield  {author} {\bibinfo {author} {\bibfnamefont {P.}~\bibnamefont
  {Niemann}}\ and\ \bibinfo {author} {\bibfnamefont {H.-W.}\ \bibnamefont
  {Hammer}},\ }\bibfield  {title} {\bibinfo {title} {Pauli-blocking effects and
  cooper triples in three-component fermi gases},\ }\href
  {https://doi.org/10.1103/PhysRevA.86.013628} {\bibfield  {journal} {\bibinfo
  {journal} {Phys. Rev. A}\ }\textbf {\bibinfo {volume} {86}},\ \bibinfo
  {pages} {013628} (\bibinfo {year} {2012})}\BibitemShut {NoStop}%
\bibitem [{\citenamefont {Kirk}\ and\ \citenamefont
  {Parish}(2017)}]{Kirk2017Phys.Rev.A96.053614}%
  \BibitemOpen
  \bibfield  {author} {\bibinfo {author} {\bibfnamefont {T.}~\bibnamefont
  {Kirk}}\ and\ \bibinfo {author} {\bibfnamefont {M.~M.}\ \bibnamefont
  {Parish}},\ }\bibfield  {title} {\bibinfo {title} {Three-body correlations in
  a two-dimensional su(3) fermi gas},\ }\href
  {https://doi.org/10.1103/PhysRevA.96.053614} {\bibfield  {journal} {\bibinfo
  {journal} {Phys. Rev. A}\ }\textbf {\bibinfo {volume} {96}},\ \bibinfo
  {pages} {053614} (\bibinfo {year} {2017})}\BibitemShut {NoStop}%
\bibitem [{\citenamefont {Tajima}\ \emph
  {et~al.}(2021{\natexlab{a}})\citenamefont {Tajima}, \citenamefont {Tsutsui},
  \citenamefont {Doi},\ and\ \citenamefont
  {Iida}}]{Tajima2021Phys.Rev.A104.053328}%
  \BibitemOpen
  \bibfield  {author} {\bibinfo {author} {\bibfnamefont {H.}~\bibnamefont
  {Tajima}}, \bibinfo {author} {\bibfnamefont {S.}~\bibnamefont {Tsutsui}},
  \bibinfo {author} {\bibfnamefont {T.~M.}\ \bibnamefont {Doi}},\ and\ \bibinfo
  {author} {\bibfnamefont {K.}~\bibnamefont {Iida}},\ }\bibfield  {title}
  {\bibinfo {title} {Three-body crossover from a Cooper triple to a bound
  trimer state in three-component Fermi gases near a triatomic resonance},\
  }\href {https://doi.org/10.1103/PhysRevA.104.053328} {\bibfield  {journal}
  {\bibinfo  {journal} {Phys. Rev. A}\ }\textbf {\bibinfo {volume} {104}},\
  \bibinfo {pages} {053328} (\bibinfo {year} {2021}{\natexlab{a}})}\BibitemShut
  {NoStop}%
\bibitem [{\citenamefont {Tajima}\ \emph {et~al.}(2022)\citenamefont {Tajima},
  \citenamefont {Tsutsui}, \citenamefont {Doi},\ and\ \citenamefont
  {Iida}}]{Tajima2022Phys.Rev.Research4.L012021}%
  \BibitemOpen
  \bibfield  {author} {\bibinfo {author} {\bibfnamefont {H.}~\bibnamefont
  {Tajima}}, \bibinfo {author} {\bibfnamefont {S.}~\bibnamefont {Tsutsui}},
  \bibinfo {author} {\bibfnamefont {T.~M.}\ \bibnamefont {Doi}},\ and\ \bibinfo
  {author} {\bibfnamefont {K.}~\bibnamefont {Iida}},\ }\bibfield  {title}
  {\bibinfo {title} {Cooper triples in attractive three-component fermions:
  Implication for hadron-quark crossover},\ }\href
  {https://doi.org/10.1103/PhysRevResearch.4.L012021} {\bibfield  {journal}
  {\bibinfo  {journal} {Phys. Rev. Research}\ }\textbf {\bibinfo {volume}
  {4}},\ \bibinfo {pages} {L012021} (\bibinfo {year} {2022})}\BibitemShut
  {NoStop}%
\bibitem [{\citenamefont {Guo}\ and\ \citenamefont
  {Tajima}(2022)}]{Guo2022Phys.Rev.A106.043310}%
  \BibitemOpen
  \bibfield  {author} {\bibinfo {author} {\bibfnamefont {Y.}~\bibnamefont
  {Guo}}\ and\ \bibinfo {author} {\bibfnamefont {H.}~\bibnamefont {Tajima}},\
  }\bibfield  {title} {\bibinfo {title} {Stability against three-body
  clustering in one-dimensional spinless $p$-wave fermions},\ }\href
  {https://doi.org/10.1103/PhysRevA.106.043310} {\bibfield  {journal} {\bibinfo
   {journal} {Phys. Rev. A}\ }\textbf {\bibinfo {volume} {106}},\ \bibinfo
  {pages} {043310} (\bibinfo {year} {2022})}\BibitemShut {NoStop}%
\bibitem [{\citenamefont {Guo}\ and\ \citenamefont
  {Tajima}(2023{\natexlab{b}})}]{Guo2023Phys.Rev.A108.043303}%
  \BibitemOpen
  \bibfield  {author} {\bibinfo {author} {\bibfnamefont {Y.}~\bibnamefont
  {Guo}}\ and\ \bibinfo {author} {\bibfnamefont {H.}~\bibnamefont {Tajima}},\
  }\bibfield  {title} {\bibinfo {title} {Cooper pairing and tripling in
  one-dimensional spinless fermions with attractive two- and three-body
  forces},\ }\href {https://doi.org/10.1103/PhysRevA.108.043303} {\bibfield
  {journal} {\bibinfo  {journal} {Phys. Rev. A}\ }\textbf {\bibinfo {volume}
  {108}},\ \bibinfo {pages} {043303} (\bibinfo {year}
  {2023}{\natexlab{b}})}\BibitemShut {NoStop}%
\bibitem [{\citenamefont {R\"opke}\ \emph {et~al.}(1998)\citenamefont
  {R\"opke}, \citenamefont {Schnell}, \citenamefont {Schuck},\ and\
  \citenamefont {Nozi\`eres}}]{Roepke1998Phys.Rev.Lett.80.3177--3180}%
  \BibitemOpen
  \bibfield  {author} {\bibinfo {author} {\bibfnamefont {G.}~\bibnamefont
  {R\"opke}}, \bibinfo {author} {\bibfnamefont {A.}~\bibnamefont {Schnell}},
  \bibinfo {author} {\bibfnamefont {P.}~\bibnamefont {Schuck}},\ and\ \bibinfo
  {author} {\bibfnamefont {P.}~\bibnamefont {Nozi\`eres}},\ }\bibfield  {title}
  {\bibinfo {title} {Four-particle condensate in strongly coupled fermion
  systems},\ }\href {https://doi.org/10.1103/PhysRevLett.80.3177} {\bibfield
  {journal} {\bibinfo  {journal} {Phys. Rev. Lett.}\ }\textbf {\bibinfo
  {volume} {80}},\ \bibinfo {pages} {3177} (\bibinfo {year}
  {1998})}\BibitemShut {NoStop}%
\bibitem [{\citenamefont {Sandulescu}\ \emph {et~al.}(2012)\citenamefont
  {Sandulescu}, \citenamefont {Negrea}, \citenamefont {Dukelsky},\ and\
  \citenamefont {Johnson}}]{Sandulescu2012Phys.Rev.C85.061303}%
  \BibitemOpen
  \bibfield  {author} {\bibinfo {author} {\bibfnamefont {N.}~\bibnamefont
  {Sandulescu}}, \bibinfo {author} {\bibfnamefont {D.}~\bibnamefont {Negrea}},
  \bibinfo {author} {\bibfnamefont {J.}~\bibnamefont {Dukelsky}},\ and\
  \bibinfo {author} {\bibfnamefont {C.~W.}\ \bibnamefont {Johnson}},\
  }\bibfield  {title} {\bibinfo {title} {Quartet condensation and isovector
  pairing correlations in $N=Z$ nuclei},\ }\href
  {https://doi.org/10.1103/PhysRevC.85.061303} {\bibfield  {journal} {\bibinfo
  {journal} {Phys. Rev. C}\ }\textbf {\bibinfo {volume} {85}},\ \bibinfo
  {pages} {061303} (\bibinfo {year} {2012})}\BibitemShut {NoStop}%
\bibitem [{\citenamefont {Baran}\ and\ \citenamefont
  {Delion}(2020)}]{Baran2020Phys.Lett.B805.135462}%
  \BibitemOpen
  \bibfield  {author} {\bibinfo {author} {\bibfnamefont {V.}~\bibnamefont
  {Baran}}\ and\ \bibinfo {author} {\bibfnamefont {D.}~\bibnamefont {Delion}},\
  }\bibfield  {title} {\bibinfo {title} {A quartet BCS-like theory},\ }\href
  {https://doi.org/https://doi.org/10.1016/j.physletb.2020.135462} {\bibfield
  {journal} {\bibinfo  {journal} {Phys. Lett. B}\ }\textbf {\bibinfo {volume}
  {805}},\ \bibinfo {pages} {135462} (\bibinfo {year} {2020})}\BibitemShut
  {NoStop}%
\bibitem [{\citenamefont {Guo}\ \emph {et~al.}(2022{\natexlab{a}})\citenamefont
  {Guo}, \citenamefont {Tajima},\ and\ \citenamefont
  {Liang}}]{Guo2022Phys.Rev.C105.024317}%
  \BibitemOpen
  \bibfield  {author} {\bibinfo {author} {\bibfnamefont {Y.}~\bibnamefont
  {Guo}}, \bibinfo {author} {\bibfnamefont {H.}~\bibnamefont {Tajima}},\ and\
  \bibinfo {author} {\bibfnamefont {H.}~\bibnamefont {Liang}},\ }\bibfield
  {title} {\bibinfo {title} {Cooper quartet correlations in infinite symmetric
  nuclear matter},\ }\href {https://doi.org/10.1103/PhysRevC.105.024317}
  {\bibfield  {journal} {\bibinfo  {journal} {Phys. Rev. C}\ }\textbf {\bibinfo
  {volume} {105}},\ \bibinfo {pages} {024317} (\bibinfo {year}
  {2022}{\natexlab{a}})}\BibitemShut {NoStop}%
\bibitem [{\citenamefont {Guo}\ \emph {et~al.}(2022{\natexlab{b}})\citenamefont
  {Guo}, \citenamefont {Tajima},\ and\ \citenamefont
  {Liang}}]{Guo2022Phys.Rev.Research4.023152}%
  \BibitemOpen
  \bibfield  {author} {\bibinfo {author} {\bibfnamefont {Y.}~\bibnamefont
  {Guo}}, \bibinfo {author} {\bibfnamefont {H.}~\bibnamefont {Tajima}},\ and\
  \bibinfo {author} {\bibfnamefont {H.}~\bibnamefont {Liang}},\ }\bibfield
  {title} {\bibinfo {title} {Biexciton-like quartet condensates in an
  electron-hole liquid},\ }\href
  {https://doi.org/10.1103/PhysRevResearch.4.023152} {\bibfield  {journal}
  {\bibinfo  {journal} {Phys. Rev. Research}\ }\textbf {\bibinfo {volume}
  {4}},\ \bibinfo {pages} {023152} (\bibinfo {year}
  {2022}{\natexlab{b}})}\BibitemShut {NoStop}%
    \bibitem [{\citenamefont {Guo}\ \emph {et~al.}(2025)\citenamefont {Guo}, \citenamefont {Naito}, \citenamefont {Tajima},\ and\ \citenamefont {Liang}}]{Guo2025Phys.Rev.C112.024310}%
  \BibitemOpen
  \bibfield  {author} {\bibinfo {author} {\bibfnamefont {Y.}~\bibnamefont {Guo}}, \bibinfo {author} {\bibfnamefont {T.}~\bibnamefont {Naito}}, \bibinfo {author} {\bibfnamefont {H.}~\bibnamefont {Tajima}},\ and\ \bibinfo {author} {\bibfnamefont {H.}~\bibnamefont {Liang}},\ }\bibfield  {title} {\bibinfo {title} {Quartet correlations near the surface of $N=Z$ nuclei},\ }\href {https://doi.org/10.1103/4rqf-5kfx} {\bibfield  {journal} {\bibinfo  {journal} {Phys. Rev. C}\ }\textbf {\bibinfo {volume} {112}},\ \bibinfo {pages} {024310} (\bibinfo {year} {2025})}\BibitemShut {NoStop}%
\bibitem [{\citenamefont {Bloch}\ \emph {et~al.}(2008)\citenamefont {Bloch},
  \citenamefont {Dalibard},\ and\ \citenamefont
  {Zwerger}}]{Bloch2008Rev.Mod.Phys.80.885--964}%
  \BibitemOpen
  \bibfield  {author} {\bibinfo {author} {\bibfnamefont {I.}~\bibnamefont
  {Bloch}}, \bibinfo {author} {\bibfnamefont {J.}~\bibnamefont {Dalibard}},\
  and\ \bibinfo {author} {\bibfnamefont {W.}~\bibnamefont {Zwerger}},\
  }\bibfield  {title} {\bibinfo {title} {Many-body physics with ultracold
  gases},\ }\href {https://doi.org/10.1103/RevModPhys.80.885} {\bibfield
  {journal} {\bibinfo  {journal} {Rev. Mod. Phys.}\ }\textbf {\bibinfo {volume}
  {80}},\ \bibinfo {pages} {885} (\bibinfo {year} {2008})}\BibitemShut
  {NoStop}%
\bibitem [{\citenamefont {Moritz}\ \emph {et~al.}(2005)\citenamefont {Moritz},
  \citenamefont {St\"oferle}, \citenamefont {G\"unter}, \citenamefont
  {K\"ohl},\ and\ \citenamefont {Esslinger}}]{PhysRevLett.94.210401}%
  \BibitemOpen
  \bibfield  {author} {\bibinfo {author} {\bibfnamefont {H.}~\bibnamefont
  {Moritz}}, \bibinfo {author} {\bibfnamefont {T.}~\bibnamefont {St\"oferle}},
  \bibinfo {author} {\bibfnamefont {K.}~\bibnamefont {G\"unter}}, \bibinfo
  {author} {\bibfnamefont {M.}~\bibnamefont {K\"ohl}},\ and\ \bibinfo {author}
  {\bibfnamefont {T.}~\bibnamefont {Esslinger}},\ }\bibfield  {title} {\bibinfo
  {title} {Confinement induced molecules in a 1d fermi gas},\ }\href
  {https://doi.org/10.1103/PhysRevLett.94.210401} {\bibfield  {journal}
  {\bibinfo  {journal} {Phys. Rev. Lett.}\ }\textbf {\bibinfo {volume} {94}},\
  \bibinfo {pages} {210401} (\bibinfo {year} {2005})}\BibitemShut {NoStop}%
\bibitem [{\citenamefont {Liao}\ \emph {et~al.}(2010)\citenamefont {Liao},
  \citenamefont {Rittner}, \citenamefont {Paprotta}, \citenamefont {Li},
  \citenamefont {Partridge}, \citenamefont {Hulet}, \citenamefont {Baur},\ and\
  \citenamefont {Mueller}}]{Liao2010Nature567}%
  \BibitemOpen
  \bibfield  {author} {\bibinfo {author} {\bibfnamefont {Y.-A.}\ \bibnamefont
  {Liao}}, \bibinfo {author} {\bibfnamefont {A.~S.~C.}\ \bibnamefont
  {Rittner}}, \bibinfo {author} {\bibfnamefont {T.}~\bibnamefont {Paprotta}},
  \bibinfo {author} {\bibfnamefont {W.}~\bibnamefont {Li}}, \bibinfo {author}
  {\bibfnamefont {G.~B.}\ \bibnamefont {Partridge}}, \bibinfo {author}
  {\bibfnamefont {R.~G.}\ \bibnamefont {Hulet}}, \bibinfo {author}
  {\bibfnamefont {S.~K.}\ \bibnamefont {Baur}},\ and\ \bibinfo {author}
  {\bibfnamefont {E.~J.}\ \bibnamefont {Mueller}},\ }\bibfield  {title}
  {\bibinfo {title} {Spin-imbalance in a one-dimensional Fermi gas},\ }\href
  {https://doi.org/10.1038/nature09393} {\bibfield  {journal} {\bibinfo
  {journal} {Nature}\ }\textbf {\bibinfo {volume} {467}},\ \bibinfo {pages}
  {567} (\bibinfo {year} {2010})}\BibitemShut {NoStop}%
\bibitem [{\citenamefont {Sowiński}\ and\ \citenamefont
  {García-March}(2019)}]{Sowiski2019}%
  \BibitemOpen
  \bibfield  {author} {\bibinfo {author} {\bibfnamefont {T.}~\bibnamefont
  {Sowiński}}\ and\ \bibinfo {author} {\bibfnamefont {M.~Á.}\ \bibnamefont
  {García-March}},\ }\bibfield  {title} {\bibinfo {title} {One-dimensional mixtures of several ultracold atoms: a review},\ }\href
  {https://doi.org/10.1088/1361-6633/ab3a80} {\bibfield  {journal} {\bibinfo
  {journal} {Rep. Prog. Phys.}\ }\textbf {\bibinfo {volume} {82}},\ \bibinfo
  {pages} {104401} (\bibinfo {year} {2019})}\BibitemShut {NoStop}%
\bibitem [{\citenamefont {Alexandrou}\ \emph {et~al.}(1989)\citenamefont
  {Alexandrou}, \citenamefont {Myczkowski},\ and\ \citenamefont
  {Negele}}]{Alexandrou1989PhysRevC.39.1076}%
  \BibitemOpen
  \bibfield  {author} {\bibinfo {author} {\bibfnamefont {C.}~\bibnamefont
  {Alexandrou}}, \bibinfo {author} {\bibfnamefont {J.}~\bibnamefont
  {Myczkowski}},\ and\ \bibinfo {author} {\bibfnamefont {J.~W.}\ \bibnamefont
  {Negele}},\ }\bibfield  {title} {\bibinfo {title} {Comparison of mean-field
  and exact monte carlo solutions of a one-dimensional nuclear model},\ }\href
  {https://doi.org/10.1103/PhysRevC.39.1076} {\bibfield  {journal} {\bibinfo
  {journal} {Phys. Rev. C}\ }\textbf {\bibinfo {volume} {39}},\ \bibinfo
  {pages} {1076} (\bibinfo {year} {1989})}\BibitemShut {NoStop}%
\bibitem [{\citenamefont {Hagino}\ \emph {et~al.}(2010)\citenamefont {Hagino},
  \citenamefont {Vitturi}, \citenamefont {P{\'{e}}rez-Bernal},\ and\
  \citenamefont {Sagawa}}]{Hagino2010}%
  \BibitemOpen
  \bibfield  {author} {\bibinfo {author} {\bibfnamefont {K.}~\bibnamefont
  {Hagino}}, \bibinfo {author} {\bibfnamefont {A.}~\bibnamefont {Vitturi}},
  \bibinfo {author} {\bibfnamefont {F.}~\bibnamefont {P{\'{e}}rez-Bernal}},\
  and\ \bibinfo {author} {\bibfnamefont {H.}~\bibnamefont {Sagawa}},\
  }\bibfield  {title} {\bibinfo {title} {Two-neutron halo nuclei in one
  dimension: dineutron correlation and breakup reaction},\ }\href
  {https://doi.org/10.1088/0954-3899/38/1/015105} {\bibfield  {journal}
  {\bibinfo  {journal} {Journal of Physics G: Nuclear and Particle Physics}\
  }\textbf {\bibinfo {volume} {38}},\ \bibinfo {pages} {015105} (\bibinfo
  {year} {2010})}\BibitemShut {NoStop}%
\bibitem [{\citenamefont {Altomare}\ and\ \citenamefont
  {Chang}(2013)}]{altomare2013one}%
  \BibitemOpen
  \bibfield  {author} {\bibinfo {author} {\bibfnamefont {F.}~\bibnamefont
  {Altomare}}\ and\ \bibinfo {author} {\bibfnamefont {A.~M.}\ \bibnamefont
  {Chang}},\ } {\bibinfo {title} {One-dimensional
  superconductivity in nanowires}}\ (\bibinfo  {publisher} {John Wiley \&
  Sons, New York},\ \bibinfo {year} {2013})\BibitemShut {NoStop}%
\bibitem [{\citenamefont {Jackson}\ \emph {et~al.}(2023)\citenamefont
  {Jackson}, \citenamefont {Dale}, \citenamefont {Maki}, \citenamefont {Xie},
  \citenamefont {Olsen}, \citenamefont {Ahmed-Braun}, \citenamefont {Zhang},\
  and\ \citenamefont {Thywissen}}]{Jackson2023Phys.Rev.X13.021013}%
  \BibitemOpen
  \bibfield  {author} {\bibinfo {author} {\bibfnamefont {K.~G.}\ \bibnamefont
  {Jackson}}, \bibinfo {author} {\bibfnamefont {C.~J.}\ \bibnamefont {Dale}},
  \bibinfo {author} {\bibfnamefont {J.}~\bibnamefont {Maki}}, \bibinfo {author}
  {\bibfnamefont {K.~G.~S.}\ \bibnamefont {Xie}}, \bibinfo {author}
  {\bibfnamefont {B.~A.}\ \bibnamefont {Olsen}}, \bibinfo {author}
  {\bibfnamefont {D.~J.~M.}\ \bibnamefont {Ahmed-Braun}}, \bibinfo {author}
  {\bibfnamefont {S.}~\bibnamefont {Zhang}},\ and\ \bibinfo {author}
  {\bibfnamefont {J.~H.}\ \bibnamefont {Thywissen}},\ }\bibfield  {title}
  {\bibinfo {title} {Emergent $s$-wave interactions between identical fermions
  in quasi-one-dimensional geometries},\ }\href
  {https://doi.org/10.1103/PhysRevX.13.021013} {\bibfield  {journal} {\bibinfo
  {journal} {Phys. Rev. X}\ }\textbf {\bibinfo {volume} {13}},\ \bibinfo
  {pages} {021013} (\bibinfo {year} {2023})}\BibitemShut {NoStop}%
\bibitem [{\citenamefont {Tajima}\ \emph
  {et~al.}(2021{\natexlab{b}})\citenamefont {Tajima}, \citenamefont {Tsutsui},
  \citenamefont {Doi},\ and\ \citenamefont
  {Iida}}]{Tajima2021PhysRevA.104.023319}%
  \BibitemOpen
  \bibfield  {author} {\bibinfo {author} {\bibfnamefont {H.}~\bibnamefont
  {Tajima}}, \bibinfo {author} {\bibfnamefont {S.}~\bibnamefont {Tsutsui}},
  \bibinfo {author} {\bibfnamefont {T.~M.}\ \bibnamefont {Doi}},\ and\ \bibinfo
  {author} {\bibfnamefont {K.}~\bibnamefont {Iida}},\ }\bibfield  {title}
  {\bibinfo {title} {Unitary $p$-wave Fermi gas in one dimension},\ }\href
  {https://doi.org/10.1103/PhysRevA.104.023319} {\bibfield  {journal} {\bibinfo
   {journal} {Phys. Rev. A}\ }\textbf {\bibinfo {volume} {104}},\ \bibinfo
  {pages} {023319} (\bibinfo {year} {2021}{\natexlab{b}})}\BibitemShut
  {NoStop}%
\end{thebibliography}


\end{CJK}
\end{document}